\documentclass[11pt,final,3p,fleqn]{elsarticle}
\usepackage{setspace} 

\usepackage{listings}
\usepackage{xcolor}

\lstdefinestyle{ansysstyle}{
  basicstyle=\ttfamily\small,
  frame=single,
  backgroundcolor=\color{gray!5},
  columns=fullflexible,
  keepspaces=true,
  showstringspaces=false,
  breaklines=true,
  captionpos=b
}

\usepackage{lmodern}

\usepackage{graphics}
\usepackage{graphicx}

\usepackage{float} 
\usepackage{adjustbox} 
\usepackage{booktabs}
\usepackage{multirow}
\usepackage{array}
\usepackage{amsmath}

\usepackage{verbatim}
\usepackage{color}
\usepackage{physics}
\usepackage{relsize}
\usepackage[dvipsnames]{xcolor}
\usepackage{array}
\usepackage{bm}
\usepackage{makecell}

\usepackage{listings}
\usepackage{algorithm}
\usepackage{setspace} 

\usepackage{algorithm}
\usepackage{algpseudocode}
\usepackage{amsmath}

\usepackage{textcomp}
\usepackage[mathscr]{euscript}

\usepackage{letltxmacro}
\LetLtxMacro{\originaleqref}{\eqref}
\renewcommand{\eqref}{Eq.~\originaleqref}
\renewcommand*{\eqref}[1]{Eq.~\originaleqref{#1}}

\usepackage{natbib}
\usepackage{hyperref}

\usepackage{doi}
\usepackage{amsmath}

\usepackage{multirow}
\usepackage{placeins} 

\usepackage[bitstream-charter]{mathdesign}
\usepackage[T1]{fontenc}

\usepackage{lscape}		
\usepackage{fancyvrb}
\usepackage{tikz}
\usepackage{tikz-3dplot}

\usetikzlibrary{arrows,chains,matrix,positioning,scopes}
\usepackage{pgfplots}
\usetikzlibrary{decorations.text}
\usepackage{tikz}
    \usetikzlibrary{patterns}
    \usetikzlibrary{arrows.meta,angles}
    \usetikzlibrary{arrows.meta,angles,quotes,calc}
    \usetikzlibrary{3d}
\usepackage{rotating}
\usepackage{graphicx}
\usepackage{pgfplots, pgfplotstable}
\usepgfplotslibrary{fillbetween}
\usepackage{nicefrac}
\usepgfplotslibrary{polar, }
\pgfplotsset{compat=1.16}

\usepackage[normalem]{ulem}
  \usepackage{lineno}
\makeatletter
\newcommand{\thickhline}{%
    \noalign {\ifnum 0=`}\fi \hrule height 1pt
    \futurelet \reserved@a \@xhline
}

\usepackage{color}

\journal{Elsevier}

\usepackage{xcolor}

\newcommand{\revA}{\color{black}}
\newcommand{\revB}{\color{black}}
\newcommand{\revC}{\color{black}}

\usepackage{tikz}
\usetikzlibrary{shapes.geometric, arrows.meta, positioning}

\usepackage{tikz}
\usetikzlibrary{arrows.meta, positioning, backgrounds, fit}

\tikzstyle{block} = [rectangle, draw=black, fill=gray!5,
                     minimum width=6.15cm, minimum height=0.6cm,
                     text centered, align=center]
\tikzstyle{note} = [rectangle, minimum height=0.8cm, text centered, text=white, font=\small, anchor=south, align=center]

\tikzstyle{arrow} = [thick,->,>=Stealth]

\begin{document}

\begin{frontmatter}

\title{
 Mesoscale FEM Model of Concrete: \\ Statistical Assessment of Inherent Stress Concentrations   \\ in Dependence on Phase Heterogeneity
}
 

\author{Jan Ma\v{s}ek$^{a,b}$, Petr Miarka$^{a,b}$}
\address{$^{a}$Institute of Physics of Materials, \\ Czech Academy of Sciences, Žižkova 22, 616 00 Brno, Czech Republic, \\  e-mail: masek@ipm.cz, miarka@ipm.cz}
\address{$^{b}$Institute of~Structural Mechanics, \\
Faculty of Civil Engineering, Brno University of~Technology, Veve\v{r}\'{i} 331/95, 602 00 Brno, Czech Republic,\\
e-mail: jan.masek1@vut.cz, petr.miarka@vut.cz}

\begin{abstract}
\begin{spacing}{1} 
Concrete heterogeneity originates from its production process, which involves bonding aggregates with a binder matrix. This study presents a mesoscale finite element model (MFEM) that offers detailed insights into the fracture process at the aggregate–cement matrix interface, focusing on one of concrete’s key properties: its mechanical response. Unlike discrete models, which often average out critical stress concentrations within the mesostructure, the MFEM approach captures detailed stress distributions, revealing localized effects crucial for understanding damage evolution.

Although computationally more demanding, the MFEM leverages modern high-performance computing (HPC) to provide a detailed description of the stress field and material damage across different phases and interfaces.

\textcolor{black}{The proposed modeling framework integrates a collision-checked aggregate generation procedure, Voronoi-based mesostructure construction, and adaptive 3D meshing, forming a reusable methodology for stress analysis in heterogeneous composites. This approach offers transparent, physically interpretable parameterization of phase properties in contrast to black-box discrete models.}

\textcolor{black}{Another methodological contribution is the statistical post-processing of stress data using histogram-based analysis across cross-sectional planes. This enables quantitative evaluation of stress concentration distributions, providing valuable insights into the mesoscale mechanical response and serving as a useful visualization tool for researchers working on heterogeneous material modeling.}

Various matrix-to-aggregate stiffness ratios are considered to evaluate the influence of material heterogeneity on the stress field.
The results are based on a statistical evaluation of stress concentrations arising from variations in material stiffness. The model is applied to investigate the impact of using recycled crushed bricks as aggregates in concrete, with particular emphasis on the stiffness mismatch between the matrix and aggregates. The study examines how this stiffness contrast affects stress distribution and ultimately influences the composite’s failure mechanisms.
\textcolor{black}{Beyond this application, the MFEM framework provides a foundation for further investigations into nonlinear fracture processes, fatigue analysis, and mechanical optimization of alternative aggregate-matrix systems.}
\end{spacing}
\end{abstract}


\end{frontmatter}

{
\small
\begin{spacing}{1} 
\noindent \textbf{Publishing information}

\noindent Mašek J., Miarka P., \emph{FEM model of concrete: Statistical assessment of inherent stress concentrations in dependence on phase heterogeneity},
Finite Elements in Analysis and Design, 2025, 252, 104442, ISSN 0168-874X 

\noindent DOI: https://doi.org/10.1016/j.finel.2025.104442.

\noindent\rule{\textwidth}{0.4pt}
\noindent{
Copyright: ©2024. This version is made available under the CC-BY 4.0 license and all right remains to authors according to CZ/EU legislation. https://creativecommons.org/licenses/by/4.0/
}
\end{spacing}
}

\clearpage

\section{Introduction}
Concrete is one of the most widely used construction materials worldwide due to its structural versatility, shape variability, and long-term durability. From a continuum mechanics perspective, concrete is a heterogeneous material, with the origin of its heterogeneity rooted in the production process, which involves binding aggregates together with a binder—typically granite aggregates combined with Portland cement \cite{nawy2000fundamentals}. In structural applications, however, concrete is usually treated as a homogeneous material, with its mechanical properties defined according to established standards \cite{eurocode2}.

\begin{figure*}[t]
    \centering
    \includegraphics[width=0.99\textwidth]{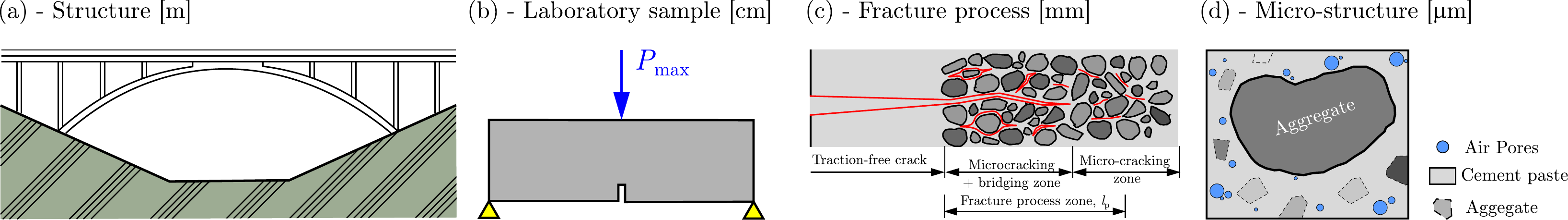}
    \caption{Scales of analysis of concrete structures.   }
    \label{fig:scales}
\end{figure*}

Nevertheless, this inherent heterogeneity becomes apparent at smaller scales, where it can significantly influence the initiation of damage. More specifically, concrete is a multi-phase material, with the number and nature of its phases depending on the scale under investigation \cite{scrivener1996percolation}. At the aggregate scale, three distinct phases can be identified: aggregates, cement paste, and air pores. Additionally, a fourth phase—the interfacial transition zone (ITZ)—represents the bonding behavior between the aggregates and the cement paste \cite{scrivener2004interfacial}. 
{\revB
The ITZ plays a significant role in damage initiation due to high porosity and bonding capability below the plain cement matrix.
ITZ is indeed an intricate region to examine mechanically, due to its small scale, heterogeneity, and complex microstructure, see the review in \cite{chen2024review}. 
}
The differences in concrete material scales are illustrated in Figure~\ref{fig:scales}.

The number of phases depends on the investigated scale. For structural applications, it is beneficial to use homogenized behavior, whereas for processes occurring at the aggregate–cement matrix level, a multi-phase material model is necessary to accurately describe the observed behavior \cite{hashin2002inverse}. This inherent heterogeneity poses a challenge for accurately calculating stresses, which are crucial for predicting future failure under both static and cyclic loading conditions.

Significant effort has been invested in developing numerical models of the mechanical response of concrete and formulating constitutive laws, which allow for more accurate predictions of stress and subsequent damage.
The most common material models are typically formulated in tensorial form within the classical continuum theory, such as the Concrete Damage Plasticity Model \cite{lee1998plastic}, the Fracture-Plastic Constitutive Model \cite{vcervenka2008three} implemented in ATENA software, microplane models \cite{caner2013microplane, zreid2018gradient}, and phase-field models \cite{yang2019x}. These models treat the material as macroscopically homogeneous without explicitly accounting for lower-scale heterogeneity.

In contrast to the constitutive material models, {\revB aggregate-resolution models (meso-scale models)} models of concrete directly involve individual heterogeneous units \cite{wriggers2006mesoscale}, which arise from the combination of a constitutive relation at a mesoscale and explicitly modeled material structure \cite{nitka2020meso}. 
A meso-scale model can provide information about the fracture processes at the level of the aggregate-to-cement matrix bond \cite{rodrigues20203d}. Thus, it can properly describe the most prominent concrete disadvantage, i.e., the strain-softening behavior as a response to tensile load \cite{scrivener2004interfacial}. Besides continuum-based models, there exist multiple other solutions such as the lattice discrete particle model (LDPM) \cite{cusatis2011lattice} or discrete finite element method (DFEM) \cite{munjiza2004combined}, where the inter-connectivity of each particle has its defined interaction.
{\revA
Phase-field models represent fractures as diffuse damage zones, simplifying crack initiation and propagation without explicit tracking. They are widely used for simulating complex fracture patterns in heterogeneous materials {\revB at the aggregate scale.} However, these models can be computationally intensive and may have limitations in accurately capturing sharp crack interfaces and fine-scale material heterogeneity \cite{nitka2018three}.
}

{\revA The discrete models, particularly LDPM, often rely on empirical contact laws and parameter calibration, limiting their interpretability and generalizability. In contrast, {\revB aggregate-resolving finite element method (MFEM) models} explicitly resolve the internal geometry and mechanical properties of each phase, providing physically interpretable fields of stress and strain and allowing direct control over material heterogeneity and microstructure geometry.}

{\revB On the other hand, the use of {\revB aggregate-resolution MFEM models} allows capturing the internal physical mechanisms of cracking as well as the macro-scale structural response. We refer the reader to models developed in 3D in \cite{zhang20193d} {\revC and 2D in \cite{zhou2018mesoscale}, respectively}.
3D mesoscale models can be used for simulating reinforced concrete’s resistance as well as ballistic penetration \cite{zhang2018numerical}.}
{\revC Also, in conjunction with cohesive elements, {\revB aggregate-resolving MFEM models} can simulate the development of microcracking within the fracture process zone, see e.g. \cite{zhou2021mesoscale}.}
{\revA The present work accompanies these concepts by developing a fully 3D, collision-checked aggregate placement and meso-geometry framework combined with adaptive meshing, constituting a reusable methodology for simulating stress distributions and damage evolution in heterogeneous composites beyond concrete alone.
Although MFEM is computationally more intensive than discrete models or continuum homogenized models, leveraging high-performance computing (HPC) and automation tools makes real-scale simulations feasible.

This combination of geometric fidelity, physical transparency, and computational practicality positions MFEM as a versatile, reusable modeling framework applicable not only to concrete but also to other heterogeneous materials where mesoscale structure strongly influences mechanical response.}

\begin{figure}[htbp]
    \centering
    \includegraphics[width=0.46\textwidth]{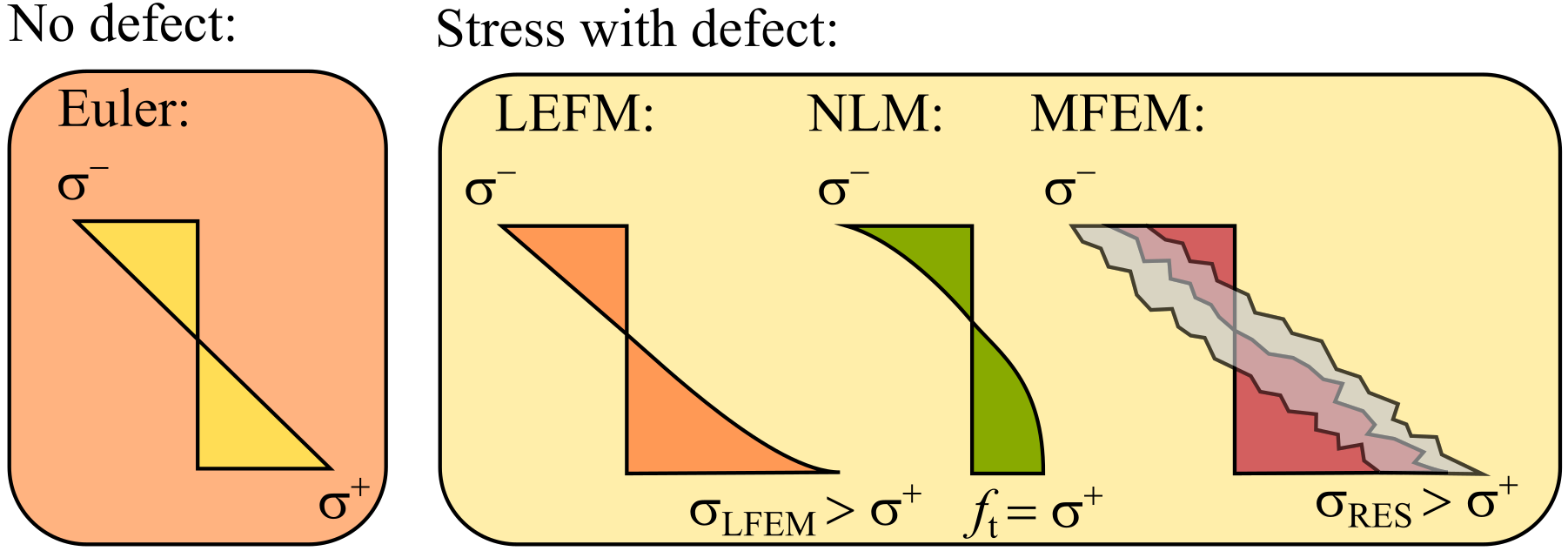}
    \caption{Illustration of normal stress caused by bending obtained by multiple calculation methods.}
    \label{fig:calcmethods}
\end{figure}

The engineering practice often desires simpler solutions than using sophisticated and often computationally expensive models. The structural design of concrete load-bearing structures typically uses well-acknowledged linear elastic beam theory, often called Euler-Bernoulli’s theory, for stress and deformation calculations, unfortunately without accounting for the high heterogeneity of concrete. The stress is then adjusted based on design recommendations in standards founded on empirical or experimental observations \cite{paris1965stress}. Figure~\ref{fig:calcmethods} compares bending stress calculated using various methods. It shows an example of how different computational approaches can simplify the stress calculation.

As shown in Figure~\ref{fig:calcmethods}, once the analyzed component contains a defect, the stress calculation must be adjusted to account for stress distribution changes along the height. Stress concentration arising from this defect can be addressed using the linear elastic fracture mechanics (LEFM) approach \cite{anderson2005fracture}, which is commonly applied in studies of metals due to strong experimental support \cite{hillerborg1976analysis} and extensive documentation in handbooks.

In the case of nonlinear material (NLM) behavior, the resulting stress at the defect location does not exceed the material’s tensile strength \cite{lee1998plastic}, and the damage is distributed (smeared) over the region of high stresses \cite{jirasek1998nonlocal}. After reaching the tensile strength \( f_t \), smeared cracking behavior is assumed, characterized by a maximum tensile strain \( \varepsilon_t \) usually related to the crack width \( w_c \). In contrast, the MFEM model accounts for local stress concentrations not only from the defect but also from the aggregate’s shape and mechanical properties.

This numerical study investigates mechanical stress distribution through 3D MFEM simulations under different loading conditions and model geometries. These include three-point bending tests on both notched and unnotched beams to induce bending stresses, as well as a compressed cylinder representing a pure compression state.

Various matrix-to-aggregate stiffness ratios are considered to evaluate the influence of material heterogeneity on the stress field.
The results are based on a statistical evaluation of stress concentrations arising from variations in material stiffness.

\section{Motivation}
\label{sec:motiovation}
The use of crushed brick as a replacement for natural aggregates in concrete offers both environmental and economic advantages. However, concerns remain regarding its mechanical performance, particularly due to the stiffness mismatch between brick aggregates and the cement matrix. Unlike granite aggregate concrete—where a uniform stiffness is often assumed based on empirical reliability—brick aggregate concrete exhibits a pronounced contrast in elastic properties. This discrepancy leads to stress concentrations at the aggregate-matrix interfaces, which can accelerate microcrack initiation and alter the overall stress distribution within the material \cite{xiao2013effects, xiao2013properties}. Understanding these variations in the stress field is crucial, as they directly influence crack propagation mechanisms and ultimately affect the structural integrity of the composite.

Recycled materials, including crushed bricks, are increasingly employed as coarse aggregates in concrete \cite{silva2016establishing}. The incorporation of aggregates with mechanical properties significantly different from traditional materials such as granite can substantially impact the mechanical response of the composite \cite{adamson2015durability}. In particular, stress concentration zones become critical, serving as initiation points for material damage and failure.
Brick aggregates present additional complexities due to their inherent variability. Their material properties can vary considerably \cite{narayanan2013properties}, with reported elastic modulus values ranging from 3.5~GPa to 34~GPa and Poisson’s ratios between 0.12 and 0.29. Moreover, the crushing process may introduce further degradation, increasing heterogeneity among individual aggregates. This variability complicates the mechanical behavior of brick aggregate concrete, necessitating a deeper investigation into its stress distribution and failure mechanisms.

{\revB
The purpose of the numerical study presented is to deliver detailed information about the internal stress distribution within concrete containing non-traditional aggregate materials. {\revA While the general effects of stiffness mismatch may have been predicted earlier, the scale of this study and the extensive statistical analysis of stress data provide a novel and unprecedented level of insight.}
In long-term, high-cycle fatigue, fracture originates from highly singular stress concentrations within a material volume otherwise well within its elastic regime.
Therefore, the crack propagation and damage evolution is not the aim of the study - the attention is devoted to the influence of using various aggregate materials on inducing stress concentrations that will eventually trigger a damage.
}

\section{Modeling approach}
{\revC Several approaches have been developed for creating and placing the aggregates within the mesostructure, notably e.g. using the Voronoi tesselation \cite{benkemoun2010failure, galindo2012breaking} or the Take-and-Place procedure \cite{zhou20173d}.}
The approach used in the present work is based on a weighted Voronoi tessellation and its adjustments for meso-structure generation informed by the material’s sieve curve.
Next, the numerical model is presented, including geometry, topology, and FE mesh generation. 

\subsection{Meso-structure topology generation}
\label{sec:mesogeneration}
\begin{figure}[htbp]
    \centering
    \includegraphics[width=0.4\textwidth]{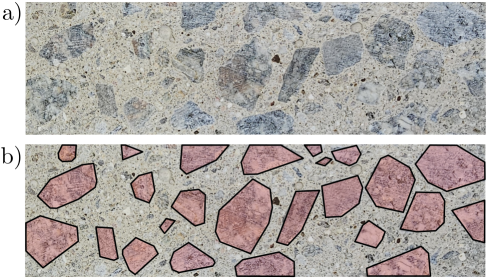}
    \caption{a) A typical concrete mesostructure, b) a possible approximation of aggregates by convex polyhedra.  }
    \label{fig:structureintro}
\end{figure}

To model the heterogeneous structure of classical concrete, we first focus on capturing the shape and distribution of the aggregates. Figure~\ref{fig:structureintro}a shows a typical concrete mesostructure containing granite aggregates embedded in a cement matrix. The shape of the aggregates depends on the material and origin of the aggregate source. However, aggregates often resemble sharp-edged convex polyhedra (e.g., crushed granite or crushed brick recyclate). Therefore, without excessive simplification, it is reasonable to utilize convex polyhedral shapes within numerical models. Figure~\ref{fig:structureintro}b illustrates how convex polyhedra can be identified in the depicted concrete mesostructure.

\begin{figure*}[t]
    \centering
    \includegraphics[width=0.95\textwidth]{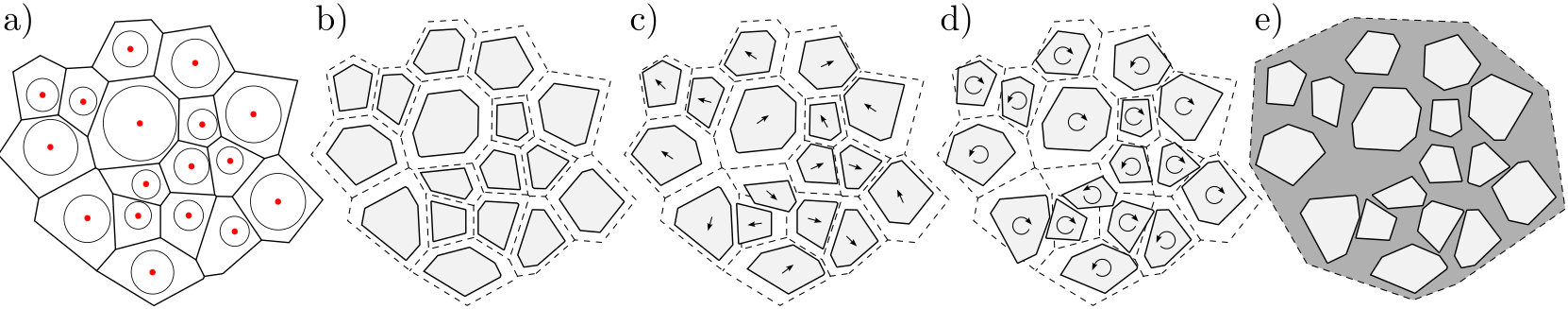}
    \caption{ An illustration of the mesostructure generation process: a) Power diagram of the input nodes, b) shrinking the cells (aggregates) towards their center of gravity to create the matrix volume, c) a random translation of the cells, d) random rotation of the cells, e) the resulting mesostructure with aggregates and cement matrix.}
    \label{fig:mesostructure}
\end{figure*}

The generation of the mesostructure begins with conducting a tessellation on a set of generating nodes. All figures in Figure~\ref{fig:mesostructure} present a 2D illustration of the mesostructure generation process. A classic Voronoi cell, \( C_i \), contains all domain points, \( \mathbf{u} \), that are closer to the respective generating node \( \mathbf{X}_i \) than to any other generating node \( \mathbf{X}_k \) in the model domain, \( \Omega \):
\begin{equation}
    C_i = \left\{ \mathbf{u} \in \Omega \;\middle|\; \forall k \neq i \quad
        \|\mathbf{u} - \mathbf{X}_i\| < \|\mathbf{u} - \mathbf{X}_k\|
    \right\}
\end{equation}

Unlike the classic Voronoi tessellation, we utilize the Power diagram variant \cite{aurenhammer1987power}. The Power diagram is a weighted version of the standard Voronoi tessellation, where each cell, \( P_i \), encompasses all domain points associated with a sphere of radius \( r_i \) around its generating node \( \mathbf{X}_i \):
\begin{equation}
    P_i = \left\{ \mathbf{u} \in \Omega \;\middle|\; \forall k \neq i \quad
        \|\mathbf{u} - \mathbf{X}_i\|^2 - r_i^2 < \|\mathbf{u} - \mathbf{X}_k\|^2 - r_k^2
    \right\}
\end{equation}
{\revA For the Power diagram generation, the DMG-$\alpha$ library is used, see \cite{szczelina2014dmg}.}

This approach enables an uneven distribution of Voronoi cell sizes that approximates a Fuller curve from \( l_{\min} \) to \( l_{\max} \), as illustrated in Figure~\ref{fig:mesostructure}a.

        The model geometry is created by randomly sequentially placing nodes into the volume domain of the intended model, \( \Omega \). We restrict the mutual distance between nodes, \( d \), to lie within the range \( d_{\min} \leq d \leq d_{\max} \). The generator starts with the maximum interparticle distance \( d = d_{\max} \) and randomly places nodes into the domain, ensuring that no two nodes are closer than \( d \). After \(10^5\) consecutive failed placement attempts, the algorithm decreases \( d \) to the next smaller value on the Fuller curve and repeats the process. The generation ends upon reaching a target domain saturation, \( S_{\mathrm{target}} \), typically set to 95\%.

To create a heterogeneous mesostructure resembling generic concrete, we shrink the cells toward their respective centers of gravity, creating intermediate zones that will eventually contain the cement matrix (see Figure~\ref{fig:mesostructure}b). The minimal distance of generating nodes and their radii during initial node placement reflect this shrinking so that the final grains have the intended size. Depending on the expected level of aggregate compaction, the shrinkage may vary between 10\% and 30\% of the cell volume.

\begin{figure*}[htbp]
        \begin{minipage}[t]{0.99\textwidth}
         \centering
        
        \includegraphics[height=3.8cm]{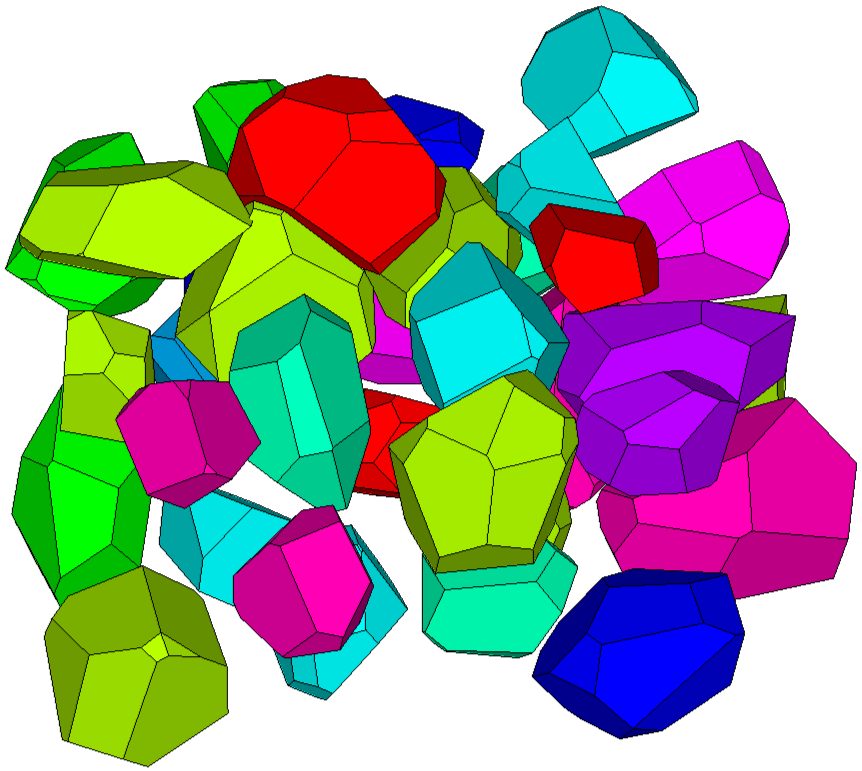}   \hspace{5mm}
        \includegraphics[height=3.8cm]{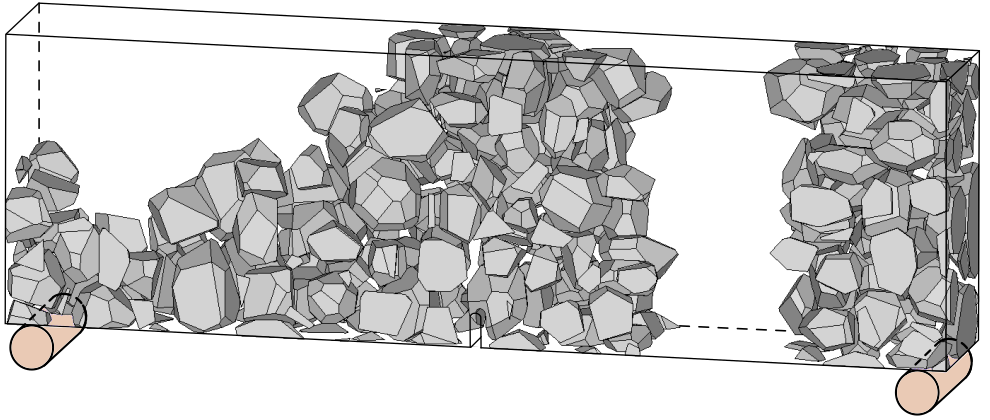} \hspace{5mm}
        \includegraphics[height=3.8cm]{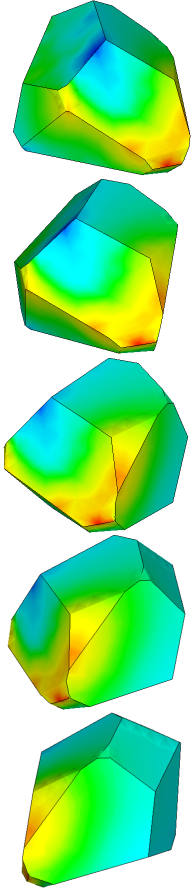}         \includegraphics[width=3.8cm, angle=90]{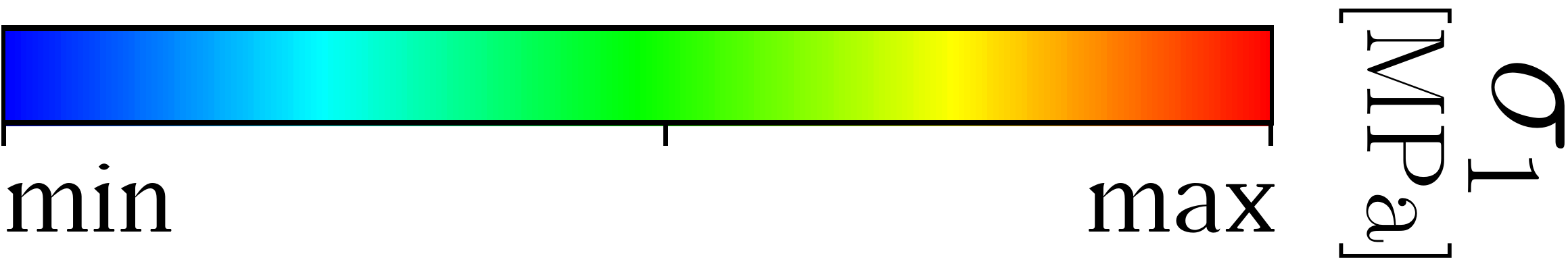} 

        \end{minipage}

        \vspace{-40mm}

         \begin{minipage}[t]{0.99\textwidth}
    \hspace{10mm}a)\hspace{42mm}b)\hspace{93mm}c)
        \end{minipage}%
        \vspace{33mm}
            
            \caption{a)\,A 3D representation of aggregates, b) example of the mesostructure within a notched three-point bending specimen, c)\,exemplary $\sigma_1$ stress field on the surface of an aggregate. }
            \label{fig:structure3d}
        \end{figure*}

To further enrich the structure, each cell is randomly shifted within the available surrounding space (see Figure~\ref{fig:mesostructure}c). Finally, to break the mutual parallelism of cell surfaces, each cell is rotated by a random 3D angle (see Figure~\ref{fig:mesostructure}d). During this process, we ensure that no two cells (grains) collide. In this way, a mesostructure resembling that of concrete is created; compare Figure~\ref{fig:mesostructure}e with Figure~\ref{fig:structureintro}. A 3D illustration of the created structure is shown in Figures~\ref{fig:structure3d}b and \ref{fig:structure3d}c.

For collision detection between grains, the Separating Axis Theorem (SAT) \cite{hyperplane_separation_theorem_wikipedia} is applied. The core idea of SAT is to identify if there exists an axis along which the projections of two polyhedra do not overlap. If such an axis is found, the polyhedra are not colliding; if no separating axis exists, a collision is detected.

The method begins by calculating the normal vectors (axes) for the faces of each polyhedron. These normals serve as potential separating axes. The vertices of the polyhedra are then projected onto these axes. If, for any axis, the projections do not overlap, the algorithm returns False, indicating no collision. If the projections overlap on all axes, the algorithm returns True, indicating a collision.
This approach is efficient and widely used for detecting collisions between convex polyhedra.
A schematic algorithm of the described mesostructure generation process is provided in Algorithm~\ref{alg:meso}.

\begin{algorithm}[htbp]
\caption{Random Sequential Node Placement and Mesostructure Generation}
\begin{algorithmic}[]
\footnotesize
    \State \textbf{Input:} Domain $\Omega$, distance range $[d_{\min}, d_{\max}]$, target saturation $S_{\text{target}}$
    \State Initialize $d \gets d_{\max}$
    \State Initialize node set $N \gets \emptyset$
    \While{Saturation $S < S_{\text{target}}$}
        \State $\text{failures} \gets 0$
        \While{$\text{failures} < 10^5$}
            \State Randomly generate node $n \in \Omega$
            \If{$\forall m \in N, \|n - m\| \geq d$}
                \State Add $n$ to $N$
                \State Update saturation $S$
                \State Reset $\text{failures} \gets 0$
            \Else
                \State Increment $\text{failures}$
            \EndIf
        \EndWhile
        \State Decrease $d$ to the next value on the Fuller curve
    \EndWhile

    \\
    \State \textbf{Step 2: Shrink Cells}
    \For{each cell $C_i$ in the structure}
        \State Compute center of gravity $G_i$
        \State Shrink $C_i$ towards $G_i$ by a factor $s \in [10\%, 30\%]$
    \EndFor
    
\\
\State \textbf{Step 3: Random Cell Shifting and Rotation}
\For{each cell $C_i$}
    \While {$C_i$ collides}
        \State Randomly shift $C_i$ within allowed bounds
        \State Rotate $C_i$ by a random 3D angle as a rigid body
        \\
        \State Collision Detection Using SAT
        \For{each already placed cell $C_j$}
            \State Face normals as potential separating axes
            \State Project vertices of $C_i$ and $C_j$ onto these axes
            \If{any axis separates the projections}
                \State No collision detected
                \State Add $C_i$ to the list of placed cells
            \Else
                \State Collision detected
                \State Reattempt random shift and rotation for $C_i$
            \EndIf
        \EndFor

    \EndWhile
\EndFor
\\
    \State \textbf{Output:} Generated mesostructure geometry
\end{algorithmic}
\label{alg:meso}
\end{algorithm}

\clearpage
{\revC 
\subsection{On aggregate granularity}

In general, Voronoi polyhedra resemble the shape of real aggregates well, however it is difficult to match the desired sieve curve.
During the generation of nodes for the weighted power tessellation, distribution of weights (radii) is defined exactly according to a Fuller curve:
\begin{equation}
P(D) = 100 \left( \frac{D}{D_{\text{max}}} \right)^{n}    
\end{equation}
where \( P(D) \) is the percentage passing for particle diameter \( D \), and \( D_{\text{max}} \) is the maximum particle diameter and \( n \) is the Fuller exponent (typically 0.5). For restrained granularity, e.g.\ 4–16 mm, a modified Fuller curve reads:
\begin{equation}
P(D) = 100 \: \frac{D^n - D_{\text{min}}^n}{D_{\text{max}}^n - D_{\text{min}}^n}    
\label{eq:corrected_fuller}
\end{equation}
where \( D_{\text{min}} \) and \( D_{\text{max}} \) are the minimum and maximum particle diameters, respectively.
The theoretical Fuller curve as described in Eq.~\ref{eq:corrected_fuller} is plotted red in the plots of Figure~\ref{fig:fuller}. 
The Fuller curve used for generation of Power tesselation radii is plotted green along with their relative histogram.

After the initial nodes are generated, the Voronoi polyhedra resulting from the tessellation no longer match the Fuller curve, and this discrepancy increases after the subsequent shrinking of the cells (see Figure~\ref{fig:mesostructure}) used for modeling the mesostructure of concrete. The Fuller curve obtained from the actual resulting 3D aggregates (more precisely the diameters of the spheres of equivalent volume) is plotted in blue along with their relative histogram.
It can be seen in Figure~\ref{fig:fuller}a that shrinking the cells by, for example, 30\% of their absolute dimension towards their center of gravity results in a significant deviation from the intended Fuller curve.
To better approximate the Fuller curve on average, one might consider increasing the limiting diameters for the Power tessellation, as shown in Figure~\ref{fig:fuller}b. Indeed, this approach yields a closer overall match.

If an even better match is desired, the cells can be shrunk more carefully by weighting their respective shrinking ratios (initially e.g., 30\%) according to their 3D volume percentile, \( p_i \), among other cells.
For instance, good results are achieved by applying a probability density function of a symmetrical  Beta distribution with parameters \(\alpha = \beta = 1.8\). The weighted shrinking coefficient for cell \( i \) then reads:
\begin{equation}
s_c(p_i) = 0.3 \cdot \frac{p_i^{\alpha - 1} (1 - p_i)^{\beta - 1}}{B(\alpha, \beta)}
\end{equation}
The weighting factor is given by the probability density function of the Beta distribution with shape parameters \(\alpha\) and \(\beta\), evaluated at the cell's volume percentile \( p_i \). 
The function \( B(\alpha, \beta) \) is the Beta function, defined as
\[
B(\alpha, \beta) = \int_0^1 t^{\alpha - 1} (1 - t)^{\beta - 1} \, dt
\]
which serves as a normalization constant to ensure that the total area under the Beta distribution probability density function equals 1. This normalization is essential for \( s_c(p_i) \) to represent a proper weighting factor.
Figure~\ref{fig:fuller}c shows that this modification leads to a solid match between the Sieve curve of the resulting 3D grains and the desired ideal Fuller curve.

In case that matching a different sieve curve is desired, the Power tesselation radii can be generated exactly as desired and a similar iterative approach of a weighted shrinking can be applied.

\begin{figure*}[t]
    \centering
    \includegraphics[width=0.99\textwidth]{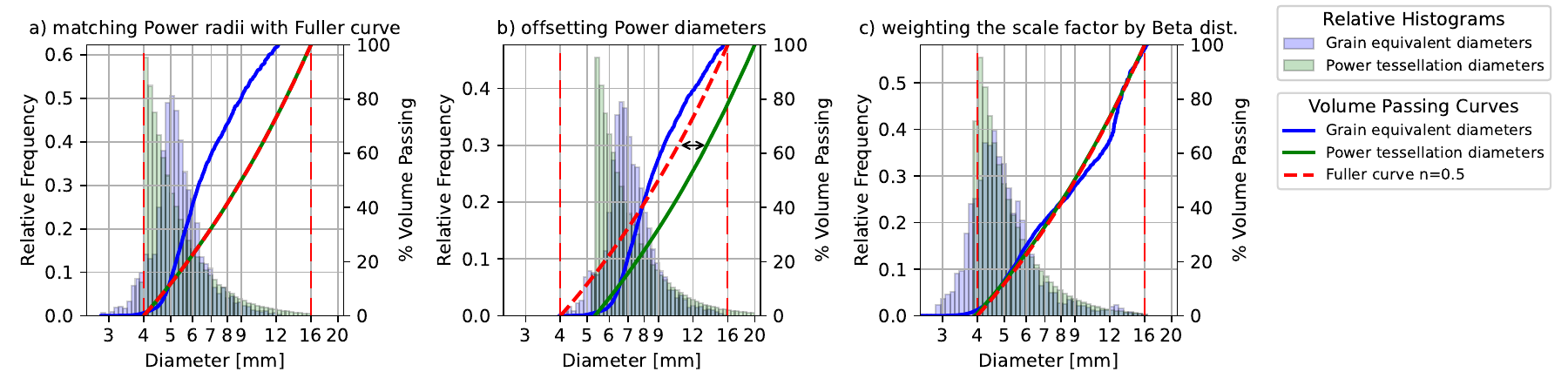}
        \caption{\revC
        Comparison of sieve curves derived from the theoretical Fuller distribution (red), the input radii for Power tessellation (green), and the actual 3D aggregate diameters after tessellation and shrinking (blue).
        (a) Uniform 30\% shrinking of Voronoi cells leads to a significant deviation from the target distribution.
        (b) Adjusting the tessellation input range reduces the mismatch.
        (c) Applying volume-percentile-weighted shrinking using e.g. a Beta distribution (\( \alpha = \beta = 1.8 \)) yields a close match to the intended Fuller curve.
    }
    \label{fig:fuller}
\end{figure*}

}

\subsection{Meshing and preprocessing}

The described mesostructure can be generated to fill the exact volume of the intended numerical model, or alternatively, the model can be extracted (cut) from a larger bulk material structure. This choice depends on whether the mesostructure near the model boundaries should reflect influences such as formwork or a specimen cut from the bulk material.

The resulting geometry and topology of the aggregate structure are imported into the ANSYS framework using the \texttt{PyMAPDL} library \cite{alexander_kaszynski_2020_4009467}, where volume objects are created in accordance with the topology.
For complex 3D topologies, mesh generation can constitute a significant portion of the total computational cost. The default ANSYS APDL mesher was found to be insufficient in terms of efficiency and performance. 
Therefore, we chose to employ the highly efficient \texttt{Gmsh} mesh generator \cite{geuzaine2009gmsh}, which offers significant improvements in both speed and flexibility. The use of the efficient Gmsh library provides a substantial preprocessing speedup compared to the proprietary ANSYS mesher. 

For volumetric meshing of the model, we use 20-node hexahedral SOLID186 elements (suitable for elastic analysis) that are degenerated into their tetrahedral forms. 
This approach is necessary to effectively mesh geometries with complex features, providing a balance between accuracy and computational feasibility. However, it is critical to carefully select the element size to avoid excessively large stress gradients within the model.

\begin{figure}[H]
    \begin{flushleft}
      \small \hspace{46mm} a) Full model \hspace{43mm} 1\,137$\times 10^3$ DOF
    \end{flushleft}\vspace{-4mm}
    \centering
    \includegraphics[width=0.49\textwidth]{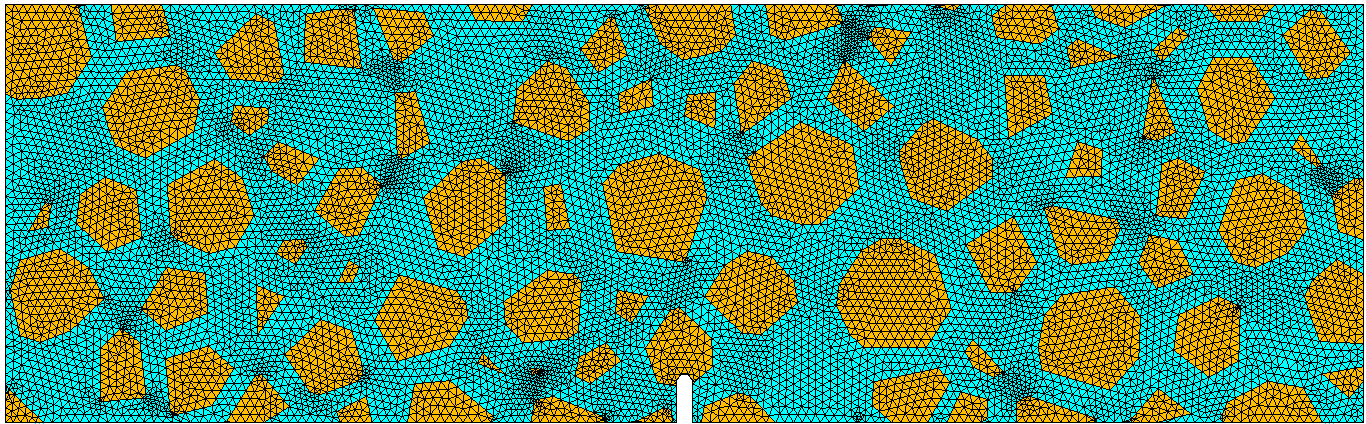}
    \vspace{-4mm}\begin{flushleft}
        \small \hspace{46mm} b) Reduced model \hspace{39mm} 228$\times 10^3$ DOF
    \end{flushleft}\vspace{-3mm}    \centering    \includegraphics[width=0.488\textwidth]{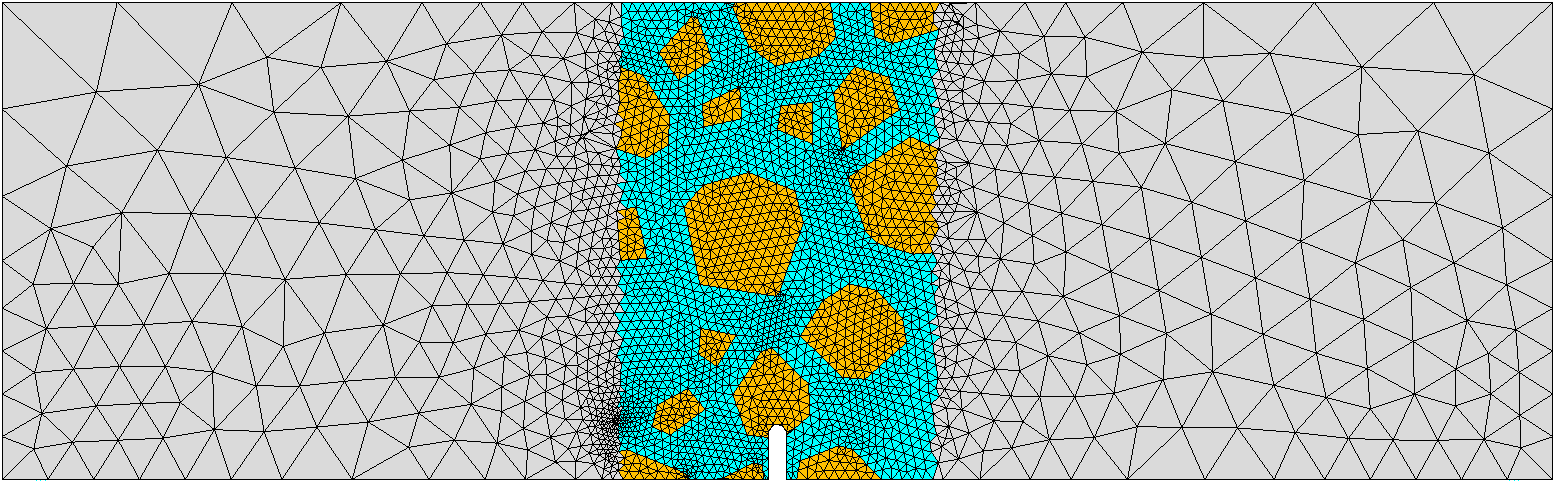}\vspace{-3mm}
    \caption{ Reducing model complexity by coarsening regions with expected low stress gradients}
    \label{fig:meshes}
\end{figure}

To optimize computational efficiency, regions of the model with expected low stress gradients are coarsened, see Figure~\ref{fig:meshes}. 
These areas exhibit minimal variations in stress distribution, allowing for mesh reduction without significant loss of accuracy.
Such an approach was verified e.g. in \cite{mavsek2023adaptive}.

In this work, without any noticeable loss of precision, computational demands were reduced by modeling and finely meshing the mesostructure only within the inner 25\% of the model width. In this fine-mesh region, the maximum element size was set to 2\,mm (see the mesh study in \ref{sec:app:meshstudy}).
{\revA
An example ANSYS-Gmsh model generating snippet is provided in \ref{sec:app:gmsh}.
}

Within the left and right outer 37\% of the model width, a coarser mesh was applied, representing a homogenized material with properties averaged between the cement matrix and aggregates. In this coarse region, the maximum element size was generally unrestricted, except for the support regions where it was limited to 6\,mm to ensure adequate resolution.

Subsequently, the assembled numerical models were executed on the computational resources of the Czech IT4Innovations National Supercomputing Center. This setup benefits from significant solution speedup due to the utilization of multiple multiprocessor computational nodes with extensive memory and storage capabilities.
{\revA
The complete modeling pipeline is shown in Figure~\ref{fig:meso_pipeline_notes}. It summarizes all steps from geometry generation to final analysis.}
\begin{figure}[ht]
\small
    \centering
    \begin{tikzpicture}

        \node (n1) [block] {Node placement};
        \node [note, text=black, fill=blue!5, minimum height=0.6cm, minimum width=6.15cm, below=0cm of n1] (note1B) {Generating nodes with mutual distances \\ varying according to a Fuller curve.};
        
        \node (n2) [block, below=3mm of note1B.south] {Power tessellation};
        \node [note, text=black, fill=blue!5, minimum height=0.6cm, minimum width=6.15cm, below=0cm of n2] (note2B) {Tessellation weighted \\ by cell generating radii.};
        
        \node (n3) [block, below=3mm of note2B.south] {Cell to aggregate transformation};
        \node [note, text=black, fill=blue!5, minimum height=0.6cm, minimum width=6.15cm, below=0cm of n3] (note3B) {Cell shrinking, translation, rotating.};
        
        \node (n4) [block, below=3mm of note3B.south] {PyMAPDL};
        \node [note, text=black, fill=teal!10, minimum height=0.6cm, minimum width=6.15cm, below=0cm of n4] (note4B) {3D model assembly};
        
        \node (n5) [block, below=3mm of note4B.south] {GMSH};
        \node [note, text=black, fill=teal!10, minimum height=0.6cm, minimum width=6.15cm, below=0cm of n5] (note5B) {Fast tetrahedra meshing.};
        
        \node (n6) [block, below=3mm of note5B.south] {PyMAPDL};
        \node [note, text=black, fill=teal!10, minimum height=0.6cm, minimum width=6.15cm, below=0cm of n6] (note6B) {Boundary conditions, load \\ stepping, solution properties.};
        
        \node (n7) [block, below=3mm of note6B.south] {IT4I HPC execution};
        \node [note, text=black, fill=red!5, minimum height=0.6cm, minimum width=6.15cm, below=0cm of n7] (note7B) {Scheduling job execution \\ on Barbora Supercomputer.};
        
        \node (n8) [block, below=3mm of note7B.south] {Post-processing};
        \node [note, text=black, fill=brown!10, minimum height=0.6cm, minimum width=6.15cm, below=0cm of n8] (note8B) {Stress field analysis via layered \\ cross-sections and histograms.};

        \node [note, rotate=90, minimum width=5.0cm, minimum height=6mm, fill=blue!50, anchor=south east] at ([xshift=-0.1cm] n1.north west) {Geometry};
        \node [note, rotate=90, minimum width=4.64cm, minimum height=6mm, fill=teal!50, anchor=south east] at ([xshift=-0.1cm] n4.north west) {Pre-processing};
        \node [note, rotate=90, minimum width=1.6cm, minimum height=6mm, fill=red!50, anchor=south east] at ([xshift=-0.1cm] n7.north west) {Solution};
        \node [note, rotate=90, minimum width=1.6cm, minimum height=6mm, fill=brown!80, anchor=south east] at ([xshift=-0.1cm] n8.north west)  {Analysis};

        \draw [arrow] (note1B.south) -- (n2.north);
        \draw [arrow] (note2B.south) -- (n3.north);
        \draw [arrow] (note3B.south) -- (n4.north);
        \draw [arrow] (note4B.south) -- (n5.north);
        \draw [arrow] (note5B.south) -- (n6.north);
        \draw [arrow] (note6B.south) -- (n7.north);
        \draw [arrow] (note7B.south) -- (n8.north);

    \end{tikzpicture}
    \caption{Stages of the MFEM modeling pipeline.}
    \label{fig:meso_pipeline_notes}
\end{figure}

\clearpage

{\revA
Table~\ref{tab:model_comparison} presents a comparison between the full and reduced finite element models in terms of computational efficiency, resource requirements, and energy consumption. The table lists the number of volume elements and degrees of freedom (DOFs) for each model, as well as the meshing time in Gmsh, the ANSYS solution time using 8 computation nodes (288 cores of Intel Cascade Lake 6240 CPU), the RAM required for the in-core solution of the FEM problem, and the total energy consumed during computation.

The full model, shown in Figure~\ref{fig:meshes}a, consists of approximately 2.04 million volume elements and 1.137 million DOFs. It requires 15 minutes for meshing, 148 GB of RAM, and 16 minutes for the solution, consuming 0.597 kWh of energy. In contrast, the reduced model, shown in Figure~\ref{fig:meshes}b, contains 406 thousand volume elements and 228 thousand DOFs, significantly reducing computational cost with a meshing time of 90 seconds, a RAM requirement of 41 GB, and an ANSYS solution time of only 3 minutes, consuming 0.107 kWh.

\begin{table}[htbp]
\smaller
\centering
\caption{FEM Model Comparison}
\label{tab:model_comparison}
\resizebox{0.42\textwidth}{!}{%
\begin{tabular}{lrr}
\toprule
& \makecell{Full model} & \makecell{Reduced model} \\
\midrule
Volume Elements & 2\,040$\times 10^3$ & 406$\times 10^3$ \\
{Degrees of Freedom} & 1\,137$\times 10^3$ & 228$\times 10^3$ \\
{Gmsh Meshing} & $15$\,min & $90$\,sec \\
{ANSYS Solution} & 10\,min & 1.5\,min \\
{RAM Required} & 148\,GB & 41\,GB \\
{Energy Used} & 0.597\,kWh & 0.107\,kWh \\
{Energy Cost} & 0.15\,EUR & 0.03\,EUR \\
\bottomrule
\end{tabular}%
}
\end{table}

Figure~\ref{fig:speedup} illustrates the trade-offs between execution time, energy consumption, and scalability. In subfigure~(a), execution time decreases significantly with increasing CPU cores, but the performance gain flattens beyond 72–108 cores, while energy consumption rises steadily.
Subfigure~(b) confirms this: stepwise speedup drops sharply with each additional node, while the energy increase ratio remains above 1 and continues to grow. All cores were utilized above 99\% throughout execution, ruling out idle hardware as a bottleneck. Each group of 36 cores corresponds to a computational node (two 18-core CPUs), so scaling introduces both intra-node and inter-node communication overheads, which progressively degrade efficiency.

Beyond 2–3 nodes, the marginal performance gain is minimal, and the energy cost continues to rise. Moreover, the HPC system is governed by SLURM-based scheduling \cite{yoo2003slurm}, where users typically wait longer in the queue to receive allocations of more nodes—even for short runs—than for allocations of fewer nodes with longer runtimes. This makes obtaining large node allocations more time-consuming and challenging in practice. In this context, it is not only more energy-efficient but also operationally advantageous to run multiple models in parallel on fewer nodes each.
This strategy improves overall solution throughput by better matching available resources and queue dynamics, rather than saturating the system with a single, inefficiently scaled job.

\begin{figure}[htpb]
\centering
\includegraphics[width=0.42\textwidth]{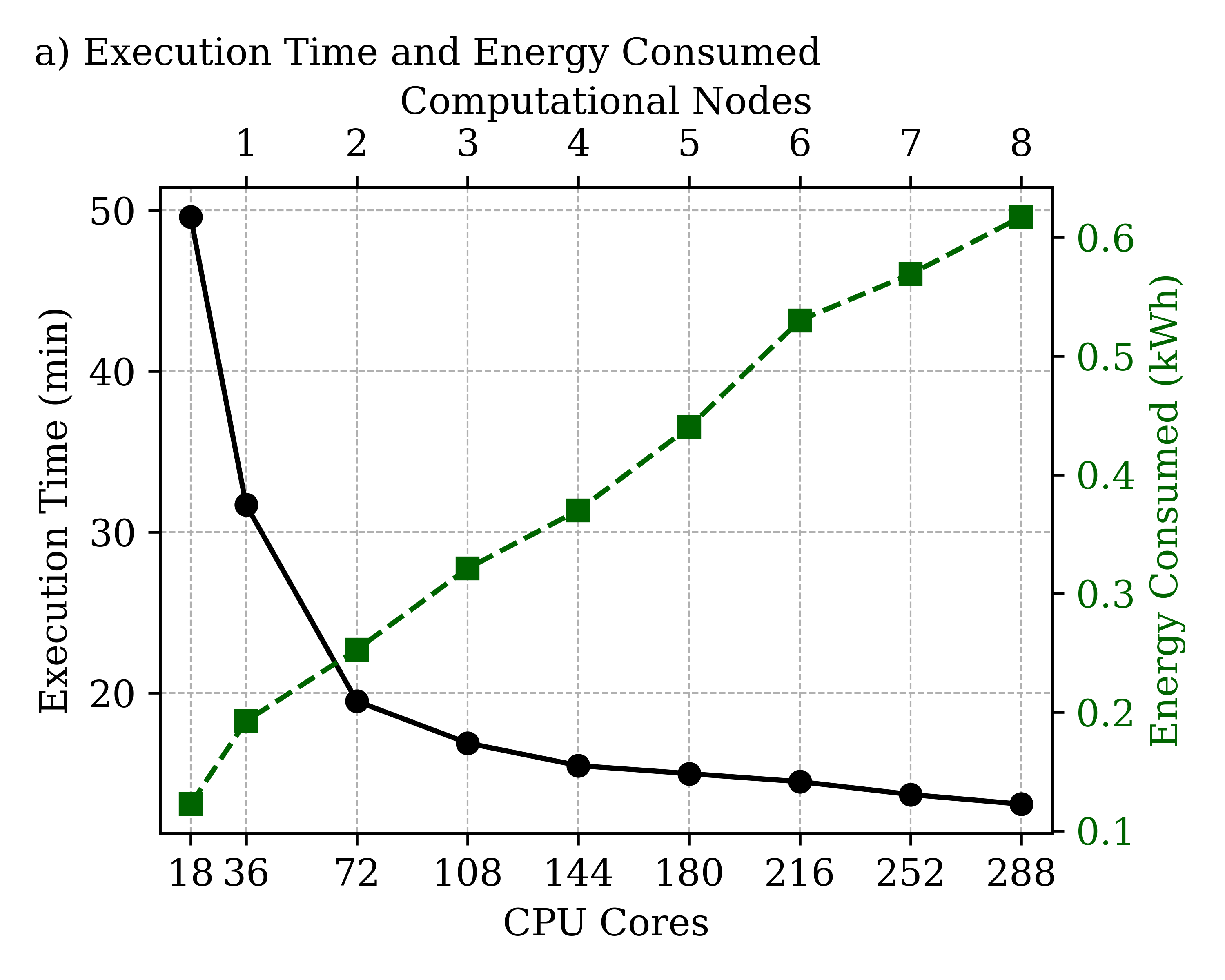}

\includegraphics[width=0.42\textwidth]{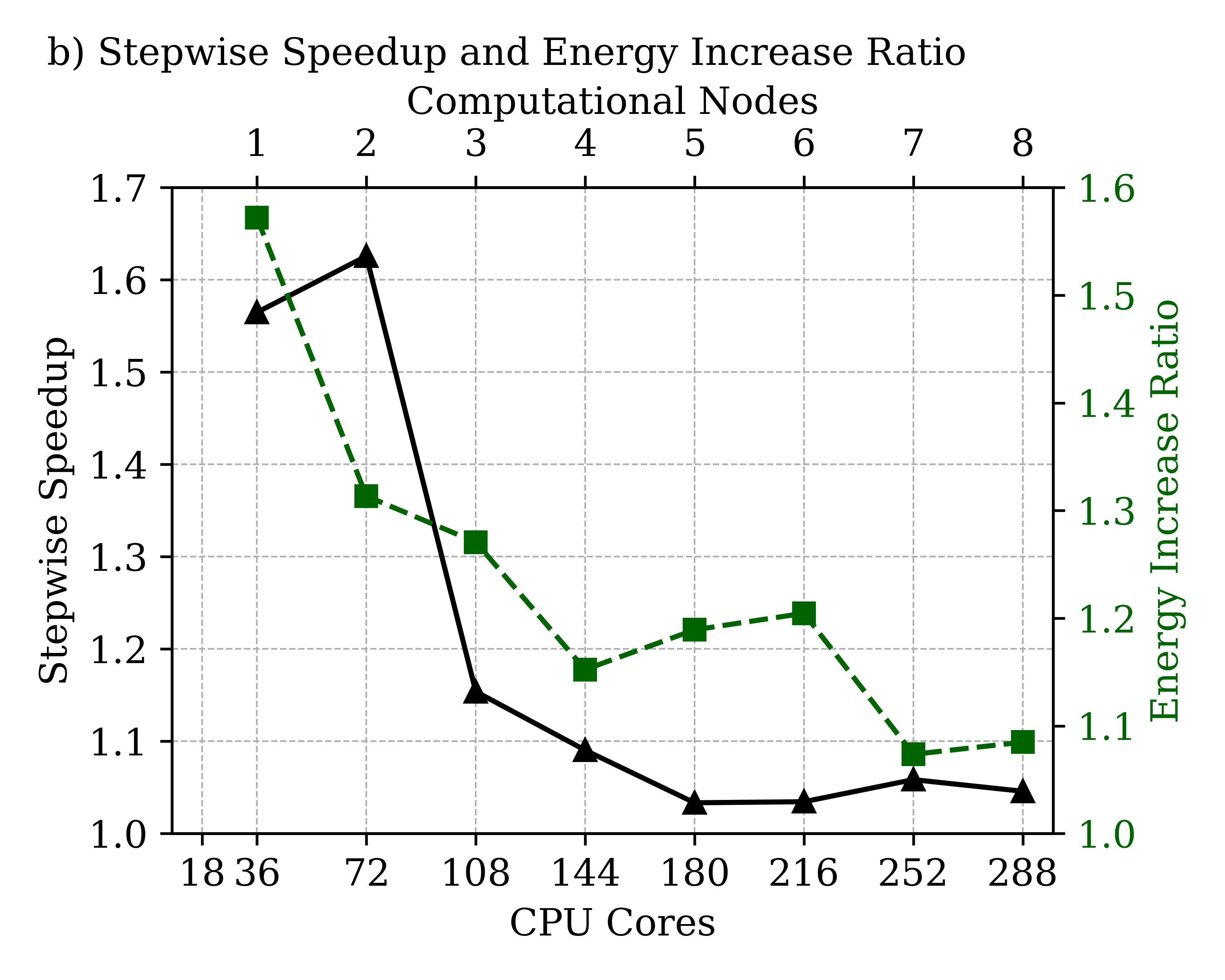}
\caption{\revA Parallelization efficiency: a)\,Execution time and energy consumed, b)\,stepwise solution speedup and stepwise energy increase ratio.}
\label{fig:speedup}
\end{figure}

}

\clearpage

\begin{figure*}[t]
    \centering
    \includegraphics[width=0.99\textwidth]{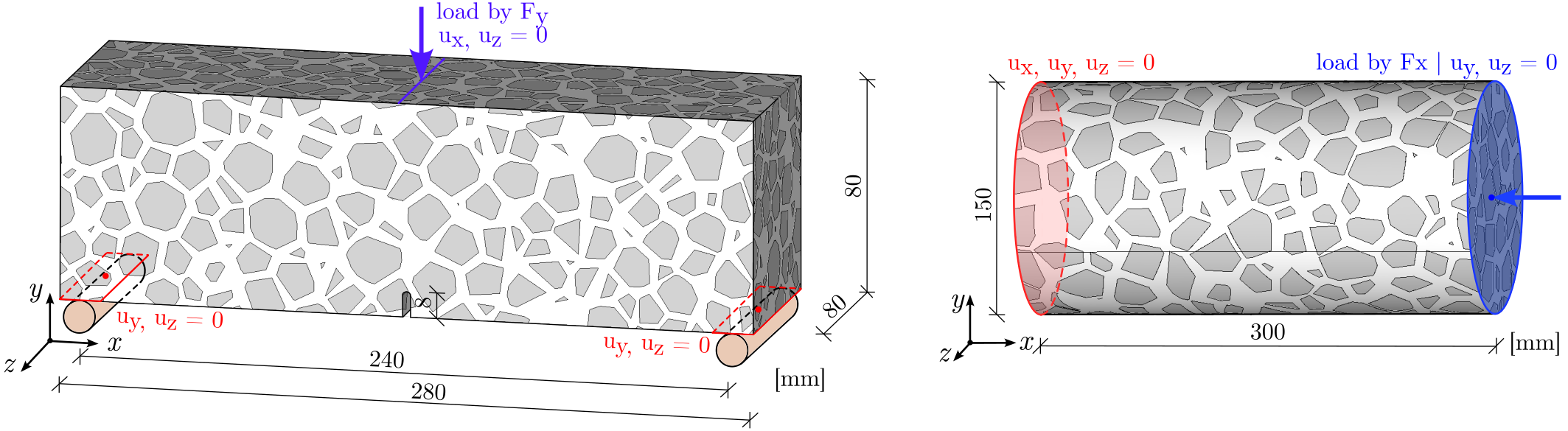}
    \caption{\revB Boundary conditions and dimensions  numerical models:  three-point bending specimen (unnotched in Sec.~\ref{sec:blunttpb}, notched in Sec.~\ref{sec:notchedtpb}) and compressed cylinder (Sec.\:\ref{sec:cylinder}).}
    \label{fig:tpbcylmodels}
\end{figure*}

\section{Numerical models}
\label{sec:numres}

In the text below, we present the geometries of the studied numerical models, including boundary conditions and analytical solutions of stress field distribution within several common concrete test specimen setups: an unnotched (blunt) three-point bending (Section~\ref{sec:blunttpb}), a notched three-point bending (Section~\ref{sec:notchedtpb}), and a compressed cylinder (Section~\ref{sec:cylinder}).

\subsection{Blunt three-point bending setup}
\label{sec:blunttpb}

The first model is a standard unnotched (blunt) three-point bending specimen. {\revB The specimen is illustrated in Figure~\ref{fig:tpbcylmodels} (consider it in the unnotched variant)}.  
The test specimen has a span \( S = 240\,\mathrm{mm} \), depth \( D = 80\,\mathrm{mm} \), thickness \( T = 80\,\mathrm{mm} \).

The boundary conditions for the supports are modeled by creating rigid regions on the beam’s bottom face (CERIG elements). Each rigid region is associated with a master node (MASS21), located at the region’s center of gravity, shown in red in Figure~\ref{fig:tpbcylmodels}. The distance between the two master support nodes corresponds exactly to the beam span \( S \).
At the supports, zero displacements are imposed in the \( y \)- and \( z \)-directions (denoted \( u_y = u_z = 0 \)), along with zero rotation about the \( X \)-axis and zero overall rotation of the model about the \( Y \)-axis.  
The loading master node is assigned a vertical force \( F_y = 1\,\mathrm{N} \) to represent a unit load; its horizontal displacements \( u_x \) and \( u_z \) are constrained to zero.

For the blunt 3PBT geometry, the uniaxial normal stress component \( \sigma_x \) is examined, as it can be directly compared to the corresponding normal stress from Euler-Bernoulli beam theory. This comparison takes into account the statistical nature of the mesostructure, as discussed in Section~\ref{sec:numanal}.

The analytical solution for the normal stress distribution in a linear homogeneous material is:
\begin{equation}
    \sigma_{x, \mathrm{LM}} = \frac{F L}{4 I} y,
\end{equation}
where \( F \) is the applied force (unitary here), \( L \) is the beam span, \( I \) is the moment of inertia of the beam cross-section, and \( y \) is the vertical distance from the neutral axis.

However, relying solely on normal stress for damage assessment can be misleading, as fracture or damage initiation often occurs under multiaxial stress states, especially when the concrete mesostructure is explicitly modeled.

\subsection{Notched three-point bending setup}
\label{sec:notchedtpb}

A standard three-point bending test setup is used, encompassing both notched and unnotched variants. The test specimen has a span \( S = 240\,\mathrm{mm} \), depth \( D = 80\,\mathrm{mm} \), thickness \( T = 80\,\mathrm{mm} \), and, for the notched variant, a notch height \( a = 0.1D \). The notch width is \(3\,\mathrm{mm}\) and its tip is rounded, as illustrated in Figure~\ref{fig:tpbcylmodels}.  
It is important to note that the stress values \( \sigma_{x_{\mathrm{LEFM}}} \) are computed assuming a sharp notch tip, so strict numerical comparison with the rounded-tip model cannot be enforced.

The boundary conditions for the top load are modeled similarly to the blunt case. A narrow rigid region is created across the top surface of the model, with a master node assigned at its center. The loading master node is assigned a vertical force \( F_y = 1\,\mathrm{N} \) representing a unit load, while its horizontal displacements \( u_x \) and \( u_z \) are constrained to zero.

The normal stress distribution within the 3D FEM model is compared to the response of a linear homogeneous material obtained from a 2D FEM counterpart model under plane strain conditions.  
This 2D model was employed to extract bending LEFM stress \( \sigma_{xx} \) along the specimen height \( W \).
To accurately capture the crack-tip singularity, the notch tip was meshed using the KSCON command, which collapses quadrilateral elements into triangular ones and moves mid-side nodes to one-quarter of the element length. The 2D mesh consisted of 8-node quadrilateral elements of type PLANE183.  
The 2D model dimensions, loading, and material parameters corresponded to those of the 3D mesoscale FEM model.
The 2D numerical solution was validated against the analytical stress intensity factor \( K_I \) from \cite{tada1973stress}, achieving an error below 1\% \cite{blason2022determination}.  
After confirming the good agreement of \( K_I \), the bending stress \( \sigma_{xx} \) was extracted for comparison.
The results demonstrate that the most frequent stress values closely align with the analytical solution, with only minor low-frequency noise around these values.

\subsection{Compression-loaded cylinder}
\label{sec:cylinder}

The second test setup is a clamped compression-loaded cylinder with height \( H = 300\,\mathrm{mm} \) and diameter \( D = 150\,\mathrm{mm} \). Rigid regions of nodes are created on both cylinder faces and assigned to their respective master nodes located at the geometric centers, as shown in Figure~\ref{fig:tpbcylmodels}.  
The master node on the loaded (top) face (blue) is constrained in all rotations and displacements except for the vertical displacement load \( u_y \).  
The master node on the supported (bottom) face (red) is fully restrained in all translations and rotations.

The normal stress distribution within the proposed FEM model is compared to the analytical solution for a linear homogeneous material:
\begin{equation}
    \sigma_{x, \mathrm{LM}} = \frac{N}{A},
\end{equation}
where \( N \) is the applied axial force (here unitary), and \( A \) is the cylinder cross-sectional area.

For a homogeneous material, the numerical model reproduces the expected mechanical response in terms of normal stress distribution as predicted by linear elastic fracture mechanics.
Experimental studies reporting stress fluctuations under compression can be found in \cite{hurley2019situ, perry1977influence, cil20143d, cil20173d}.

\clearpage
\section{Numerical results and discussion}

The sections below detail the methodology used for numerical data analysis of the studied geometries. Special attention is given to the extraction of stress data from the simulation results for each geometry. We then present the local stress distribution within aggregates derived from the internal stress state. This is followed by an in-depth analysis of the influence of stiffness ratio on the stress field, including a statistical evaluation of this complex problem.

\subsection{Numerical data analysis}
\label{sec:numanal}

To analyze the stress field in the models, we extract stress values from multiple parallel cross-sections along the specimen.  
The cross-sections are positioned sufficiently close relative to the specimen’s scale to cover volumes experiencing similar mechanical load states.  
This allows us to treat the extracted stress values as representative of a single cross-section of finite thickness. 
{\revA
For it might prove useful for a future usage, Listing~1 presents the APDL script employed to extract the specified stress values and node coordinates for each cross-section plane. 
}
{\small
\begin{lstlisting}[style=ansysstyle,caption={ANSYS APDL Script: Surface stress extraction},label={lst:ansys_export}]
! Offsets the working plane to specific x-coordinate  
WPOFFS,0.140,,                
! Rotates the working plane by 90 degrees about third rotation about Y axis so the XY WP plane becomes the global YZ plane  
WPROTA,,,90                   
! Select elements by material  
ESEL,S,MAT,,MATERIAL          
! Create a surface called surfName. Refinement level 2 
SUCR,SURFNAME,CPLANE,2        
! Map results onto selected surface  
SUMAP,RN,S,X                  
! Plot result data on selected surface  
SUPL,SURFNAME,RN,0            
! Moves surface geometry and mapped results to an array parameter  
SUGET,SURFNAME,RN,CSECTION,1    
! Retrieves array dimensions from the csection array and stores them into scalar parameters  
*GET, ARRSIZE1, PARM, CSECTION, DIM,1  
*GET, ARRSIZE2, PARM, CSECTION, DIM,2  
*GET, ARRSIZE3, PARM, CSECTION, DIM,3  
! Extracts x, y, z coordinates from csection array into a new array surf_OUT with dimensions arrSize1 by (arrSize2 - 4)  
*DIM,SURF_OUT,ARRAY,ARRSIZE1,ARRSIZE2-4  
*VFUN,SURF_OUT(1,1),COPY,CSECTION(1,1)  
*VFUN,SURF_OUT(1,2),COPY,CSECTION(1,2)  
*VFUN,SURF_OUT(1,3),COPY,CSECTION(1,3)  
! Copy eighth component (SX stress) from csection array to surf_OUT  
*VFUN,SURF_OUT(1,4),COPY,CSECTION(1,8)  
! Write the surf_OUT array data to text file named by %FILENAME1% at path %PATH1%  
*MWRITE,SURF_OUT(1,1,1),'%FILENAME1%',TXT,'%PATH1%'  
! Format specifier for output file: 7 exponential numbers per line with width 20, 8 decimals  
(7E20.8)
\end{lstlisting}
}

\begin{figure}[htpb]
\centering
\includegraphics[width=0.48\textwidth]{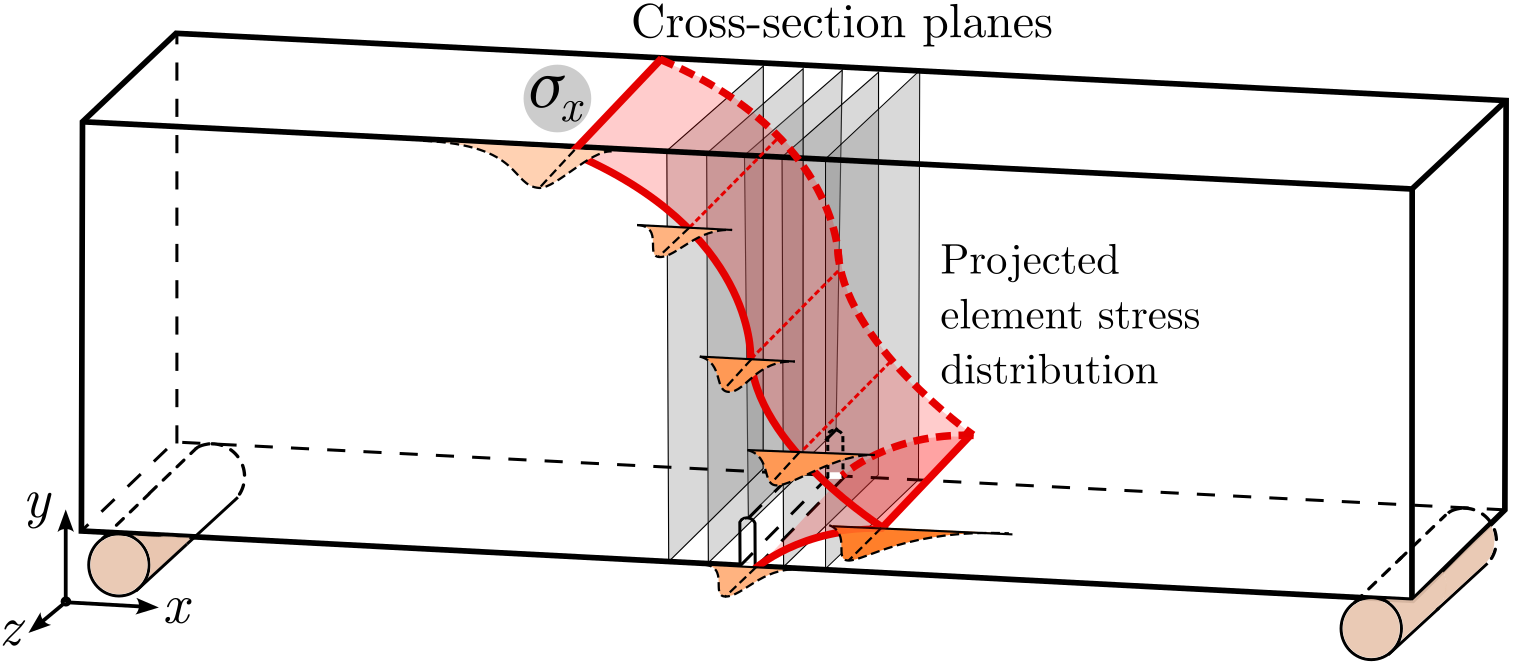}
\caption{Cross-sectional planes used for statistical data analysis.}
\label{fig:stress}
\end{figure}

\begin{figure}[htpb]

\hspace{52mm}
$\sigma_{\mathrm{x, agg}}$
\hspace{18mm}
$\sigma_{\mathrm{x, mx}}$
\hspace{21mm}
$\sigma_{\mathrm{x}}$

\vspace{0mm}
\hspace{40mm}
 \adjustbox{valign=c}{\includegraphics[height=3cm]{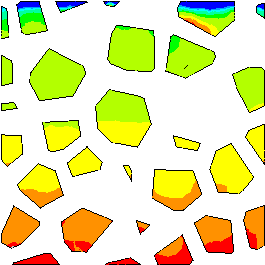}}
    \adjustbox{valign=c}{\includegraphics[height=3cm]{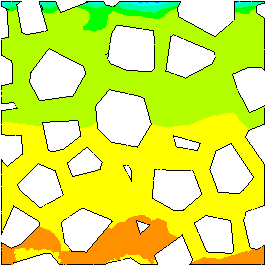}}
    \adjustbox{valign=c}{\includegraphics[height=3cm]{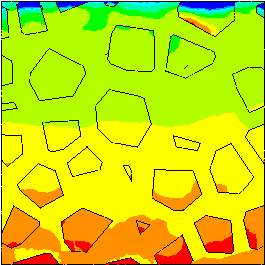}}
    \adjustbox{valign=c}{\includegraphics[height=3.1cm]{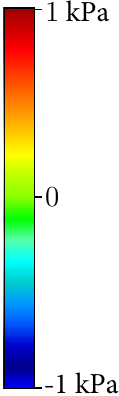}}

\caption{Normal stress across a 2D cross-section within the aggregate material ($\sigma_{\mathrm{x, agg}}$), matrix material ($\sigma_{\mathrm{x, mx}}$) and full cross-section ($\sigma_{\mathrm{x}}$).}
\label{fig:stress_2d}
\end{figure}

Without significant loss of information, we further condense the data by dividing the cross-section height into many thin layers, over which histograms of stress values are constructed.  
By assembling these histograms at their respective vertical positions, we obtain a comprehensive summary of the stress distribution across the investigated volume of the model.
This approach is illustrated in Figure~\ref{fig:stress} for a three-point bending test specimen.  
Figure~\ref{fig:stress_2d} shows an example of the normal stress field across such a cross-section, highlighting differences between aggregate and matrix phases with unequal Young’s moduli.

By this evaluation method, no information is lost or averaged out; the visualization preserves details on stress extremes as well as the full distribution of stresses.

The results presented below were acquired within 10 equidistant cross-sections   spread across the notch width, extracted from 50 numerical models with unique aggregate structures.

\subsection{Influence of Stiffness Contrast on Local Stress Distribution}

\begin{figure}[htbp]
    \centering
    \includegraphics[width=0.3\textwidth]{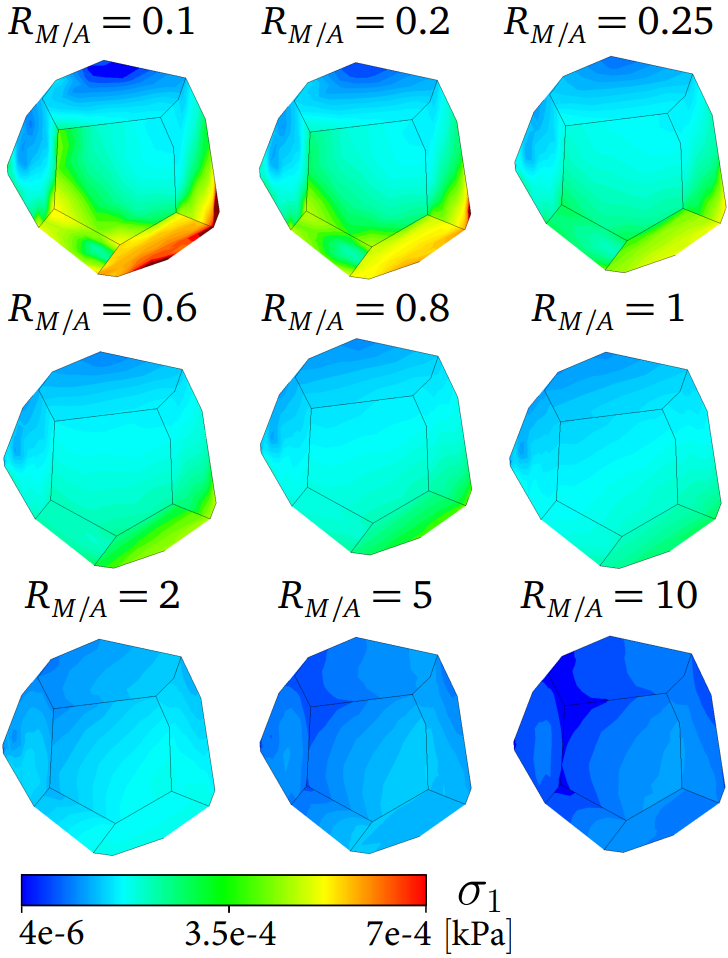}
    \caption{Development of the principal stress $\sigma_1$ distribution on a grain surface as influenced by stiffness heterogeneity.}
    \label{fig:sigma1_distribution_S1}
\end{figure}

\begin{figure}[htbp]
    \centering
    \includegraphics[width=0.3\textwidth]{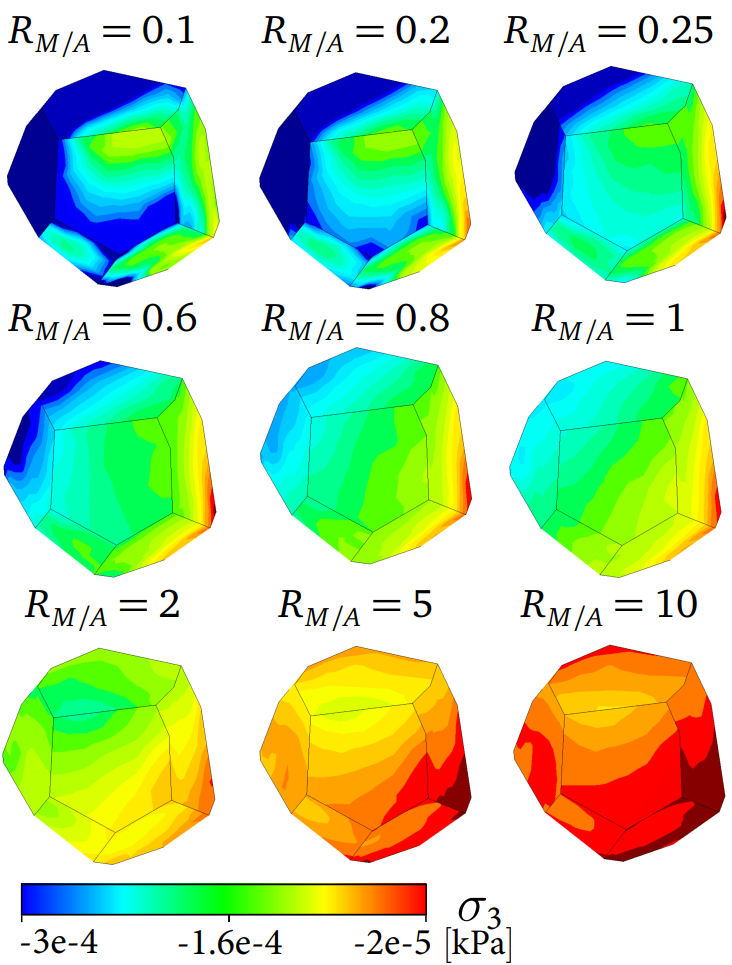}
     \caption{Development of the principal stress $\sigma_3$ distribution on a grain surface as influenced by stiffness heterogeneity.}
    \label{fig:sigma1_distribution_S3}
\end{figure}

To assess the effect of mesoscale stiffness heterogeneity on the internal stress state of concrete, we conducted an elastic finite element analysis focusing on a single polyhedral aggregate embedded within a continuous cementitious matrix.  
The first principal stress, \(\sigma_1\), was evaluated under uniaxial loading conditions for varying stiffness ratios defined as \( R_{M/A} = E_{\mathrm{mx}} / E_{\mathrm{agg}} \), ranging from 0.1 to 10.  
As illustrated in Figures~\ref{fig:sigma1_distribution_S1} and \ref{fig:sigma1_distribution_S3}, the stress field within the inclusion is highly sensitive to \( R_{M/A} \).

{\revB For softer aggregates (\( R_{M/A} > 1 \))}, the aggregate remains nearly stress-free, with the matrix accommodating the majority of the external load.  
{\revB Conversely, for softer matrix \( (R_{M/A} < 1) \), the aggregate attracts and concentrates both tensile and compressive stresses, with pronounced localization of stress near sharp geometric features such as corners and edges, see Figure~\ref{fig:sigma1_distribution_S1} for $\sigma_1$ and Figure~\ref{fig:sigma1_distribution_S3} for $\sigma_3$ principal stress, respectively.}

This behavior aligns qualitatively with classical inclusion theory—specifically Eshelby’s solution for ellipsoidal inclusions in infinite media~\cite{eshelby1957}. According to Eshelby, a homogeneous ellipsoidal inclusion embedded in a linearly elastic infinite matrix under uniform remote stress experiences a uniform internal stress and strain field, which can be expressed analytically.  
However, our case involves a faceted (polyhedral) inclusion with non-ellipsoidal geometry, where Eshelby’s assumptions no longer hold.  
In such cases, the internal stress field becomes non-uniform and exhibits singularities at geometric discontinuities.  
Analytical and numerical extensions, such as those presented in \cite{rodin1996}, confirm that polyhedral inclusions produce more complex stress fields, with stress intensification driven both by stiffness mismatch and inclusion geometry.

These findings highlight the pivotal role of stiffness contrast and morphological features in governing local stress distributions, which are central to the initiation of microcracks and the subsequent fracture evolution in quasi-brittle heterogeneous materials like concrete.

\clearpage
\subsection{Stress field dependence on aggregate to matrix stiffness ratio}
\label{sec:stiffnessratiostress}
In what follows, we investigate the influence of increasing stiffness contrast between the aggregate and matrix material phases.  
The aim is to provide insight into how the stress distribution within the material evolves as the heterogeneity of material properties between the matrix and aggregate phases increases.

This investigation is particularly relevant for the utilization of alternative aggregate materials, such as recycled crushed bricks (see Section~\ref{sec:motiovation}).  
The presence of inclusions significantly softer (lower Young’s modulus) than granite, or even softer than the cement matrix, can lead to substantial increases in stress concentrations within the composite.  
Such elevated stress concentrations are expected to cause earlier crack initiation during the service life compared to traditional granite aggregates.  
We believe that these numerical results contribute to a better understanding of concrete’s inherent heterogeneity and the resulting local stress concentrations.

We analyze the evolution of normal stress distribution differences between a homogeneous material and increasingly heterogeneous composites.

As noted in the introduction, the elastic modulus of aggregates can vary widely—from about 3.5\,GPa (e.g., crushed bricks) up to 75\,GPa or more (e.g., crushed granite).  
The cement matrix modulus typically ranges between 10\,GPa and 30\,GPa \cite{wei2020role}.  
This results in stiffness ratios \( R_{M/A} \) spanning roughly from 0.1 to 10. More elastic properties of various materials can be found \cite{bass1995elasticity}.
For this study, the Poisson’s ratio is fixed at \(\nu = 0.2\) for both materials.  
The aggregate structure is generated to resemble a realistic distribution of aggregate sizes between 4 and 16\,mm.

The MFEM stress values are displayed as relative histograms across a fine array of layers along the specimen height.  
In each layer, dark red indicates the most frequent stress values, with frequency decreasing towards dark blue.  
For homogeneous models (\(R_{M/A}=1\)), the most frequent stress values closely match the analytical reference solution, with scattered noise of negligible frequency around these values.  
Only near the top surface of the TPB models do compressive stress concentrations appear due to the loading boundary condition.

When stiffnesses are comparable (\(R_{M/A} \approx 1\)), strains in matrix and aggregates are compatible, resulting in a smooth stress distribution.  
However, as the heterogeneity becomes more pronounced in either direction, the normal stress distributions within matrix and aggregates diverge.  

For a softer matrix relative to aggregates (\(R_{M/A} < 1\)), stress peaks within the matrix tend to smear out because the matrix deforms more easily while transferring load between aggregates.  
Nonetheless, the matrix remains responsible for overall equilibrium, so its stress cannot approach zero.

Conversely, for softer aggregates relative to the matrix (\(R_{M/A} > 1\)), the stiffer matrix attracts more stress due to its higher deformation resistance.  
As aggregate stiffness decreases, they lose the ability to effectively transfer stress across the composite, forcing the matrix to carry a larger portion of the load.  
In the limiting case, the composite behaves like a cement matrix with void inclusions.

\begin{figure*}[t]
\centering
\includegraphics[height=0.46\textheight]{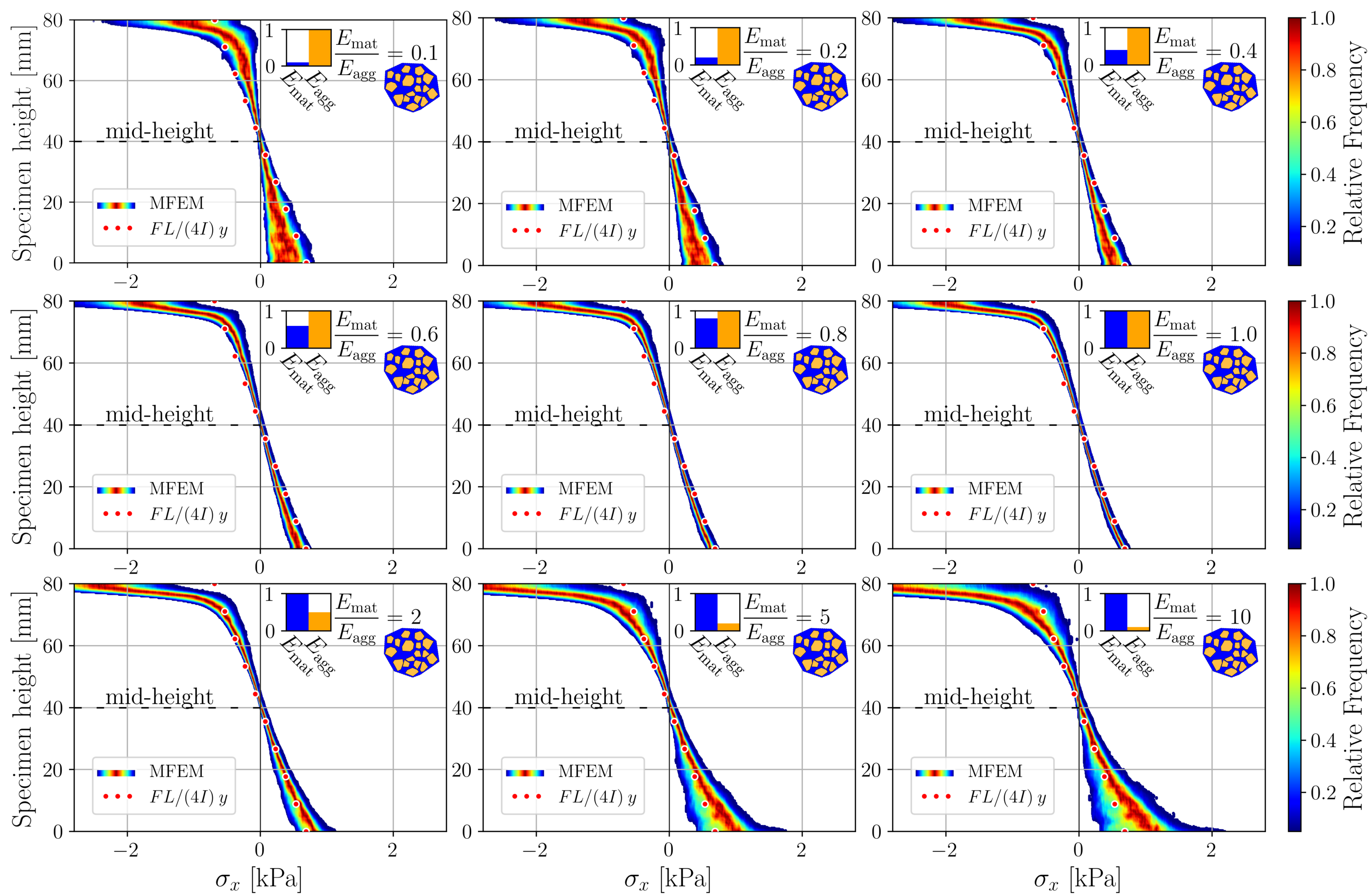}
\vspace{-3mm}
\caption{Blunt TPB: matrix normal stress within the notch region in dependence on the heterogeneity parameter, $R_{M/A}$. Colors show relative stress frequency per horizontal layer (red = most frequent).}
\label{fig:bluntmergedstressesmatrix}
\vspace{-0mm}
\includegraphics[height=0.46\textheight]{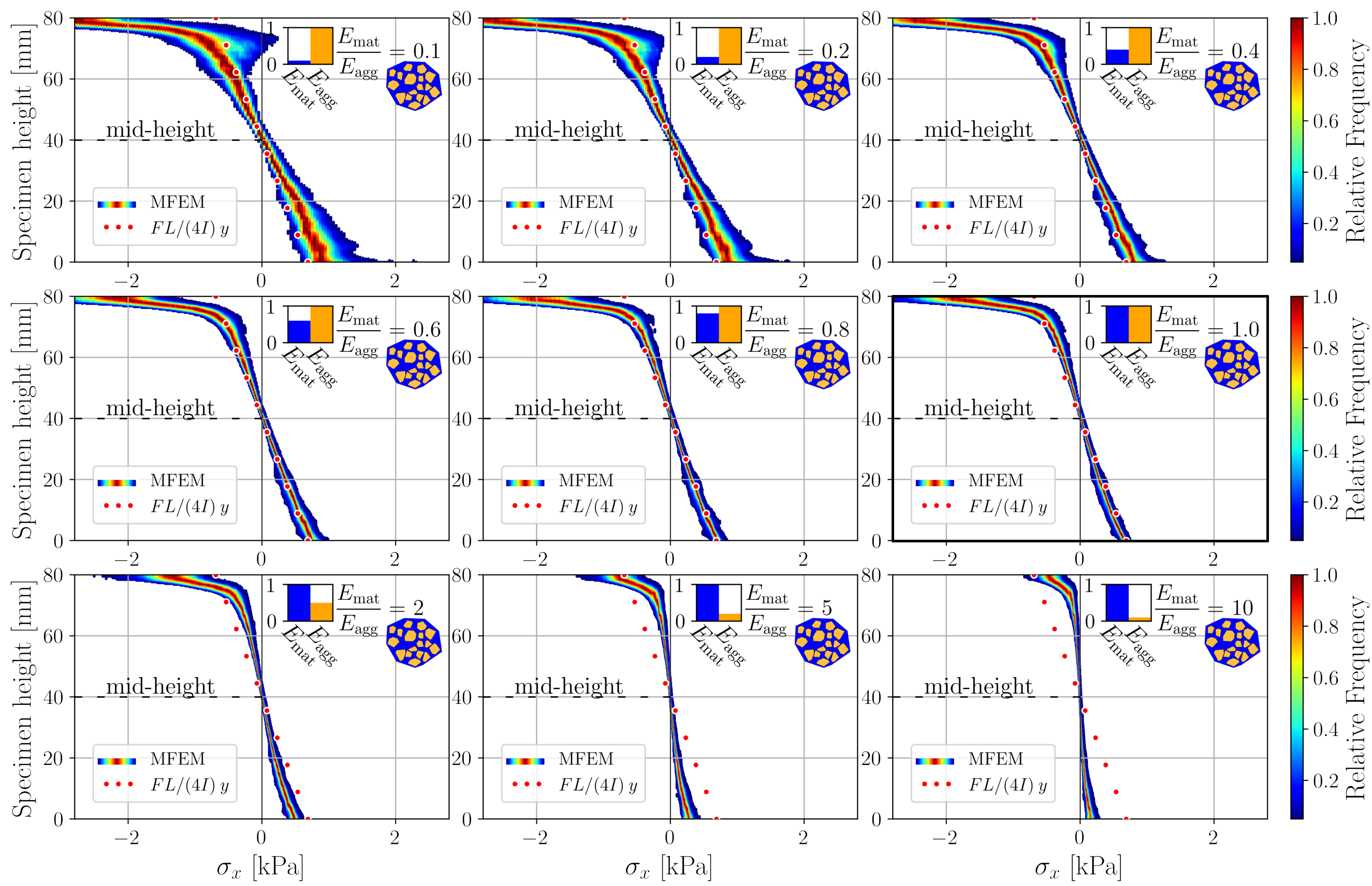}
\vspace{-3mm}
\caption{Blunt TPB: aggregates normal stress within the notch region in dependence on the heterogeneity parameter, $R_{M/A}$. Colors show relative stress frequency per horizontal layer (red = most frequent).}
\label{fig:bluntmergedstressesagg}
\end{figure*}

\begin{figure*}[t]
\centering
\includegraphics[height=0.43\textheight]{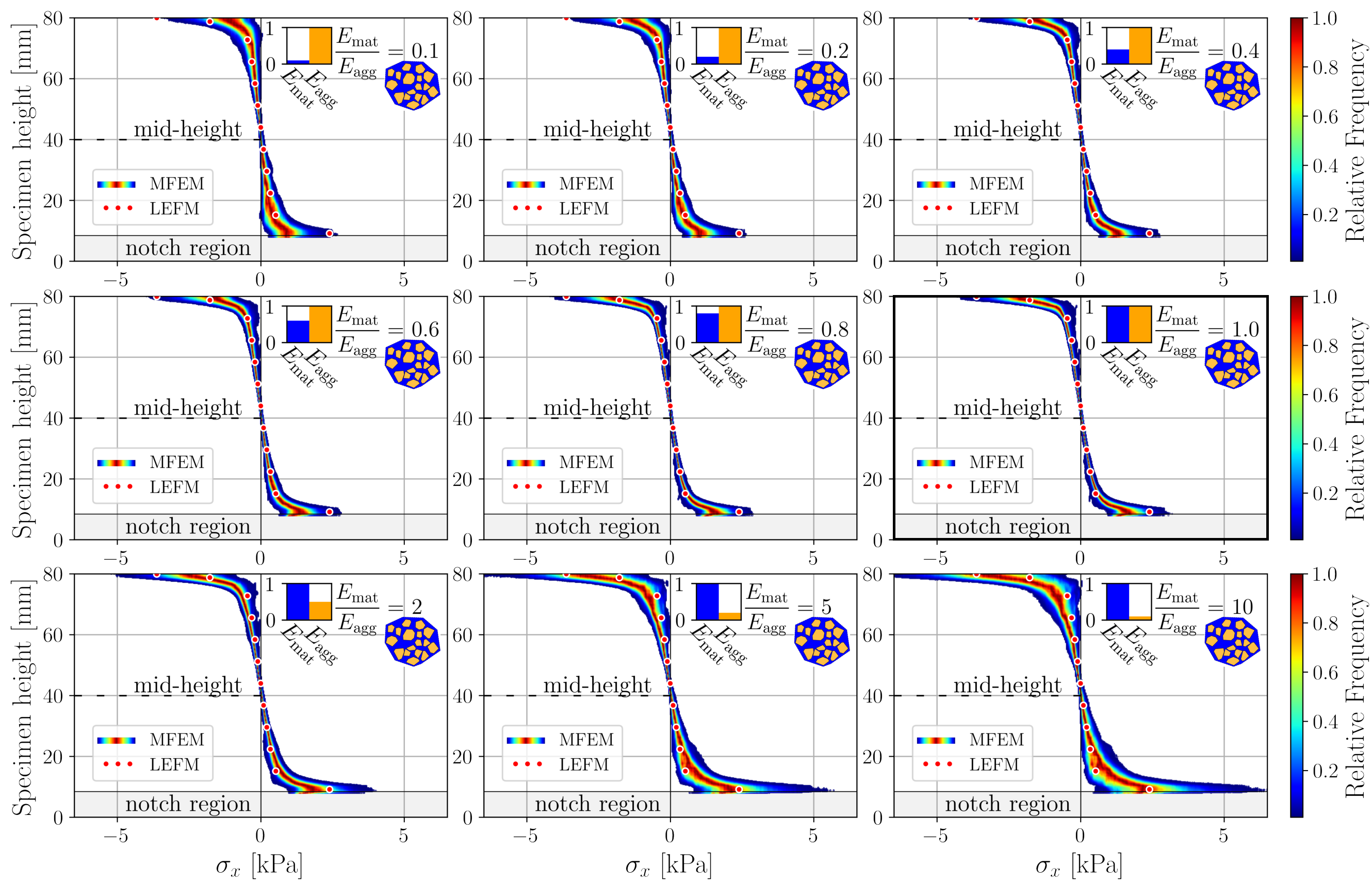}
\vspace{-2mm}
\caption{Notched TPB: matrix normal stress within the notch region in dependence on the heterogeneity parameter, $R_{M/A}$. Colors show relative stress frequency per horizontal layer (red = most frequent).}
\label{fig:mergedstressesmatrix}
\vspace{0mm}
\includegraphics[height=0.43\textheight]{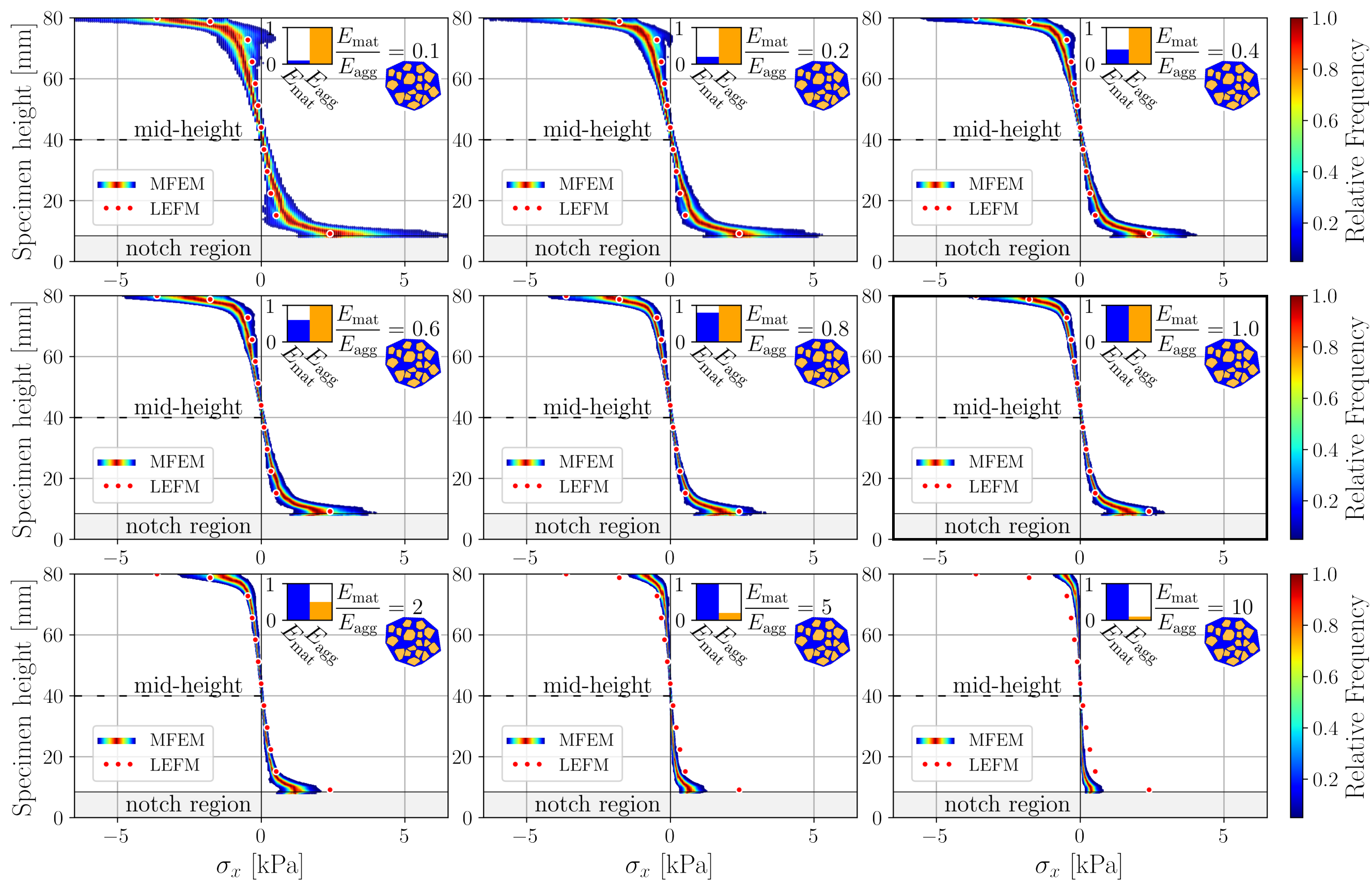}
\vspace{-3mm}
\caption{Notched TPB: aggregates normal stress within the notch region in dependence on the heterogeneity parameter, $R_{M/A}$. Colors show relative stress frequency per horizontal layer (red = most frequent).}
\label{fig:mergedstressesagg}
\end{figure*}

\begin{figure*}[t]
\centering
\includegraphics[height=0.43\textheight]{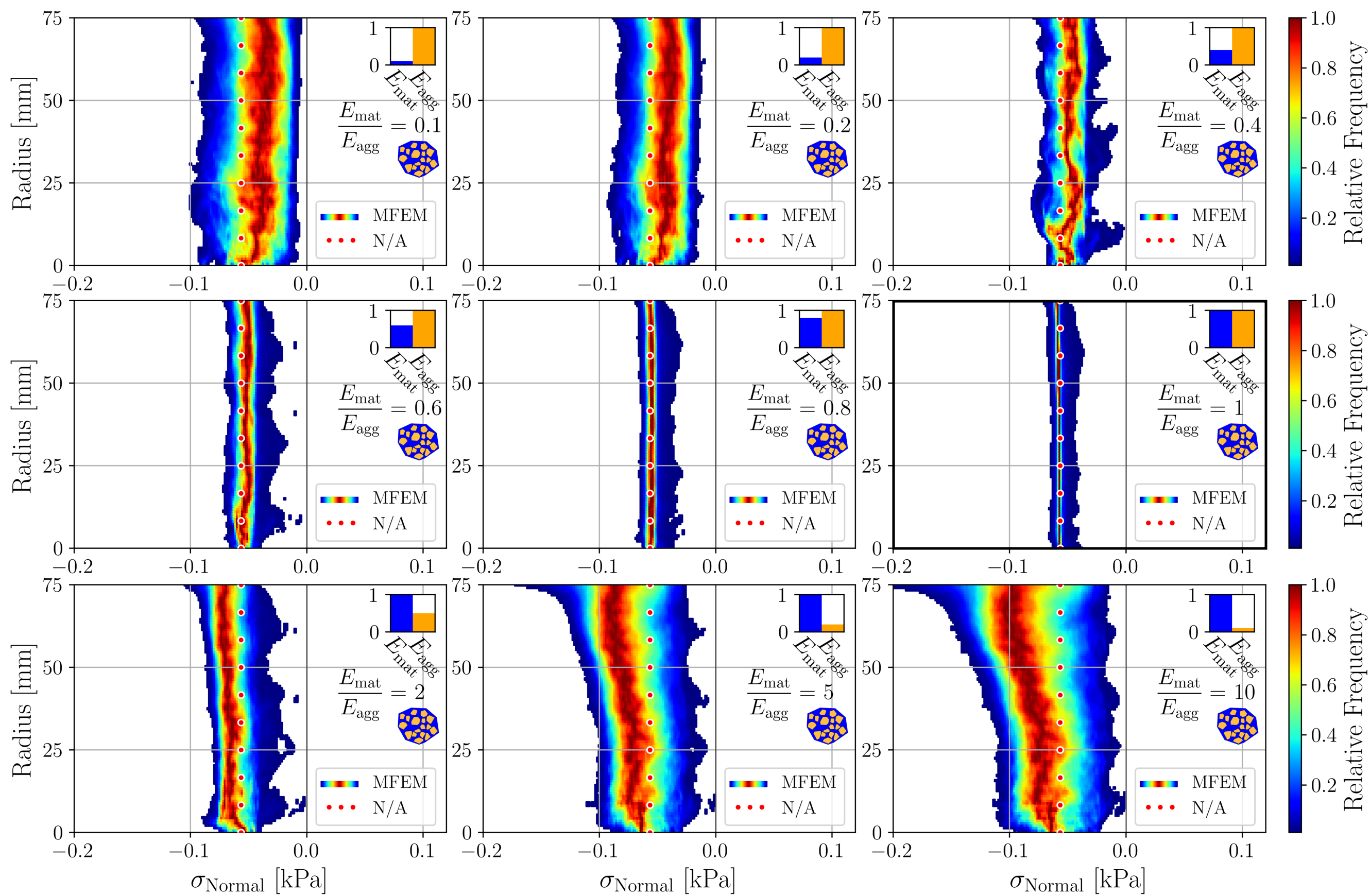}
\vspace{-3mm}
\caption{Cylinder compression: matrix normal stress at midspan region in dependence on the heterogeneity parameter, $R_{M/A}$. Colors show relative stress frequency per horizontal layer (red = most frequent).}
\label{fig:mergedstressesmatrixcyl}
\vspace{0mm}
\includegraphics[height=0.43\textheight]{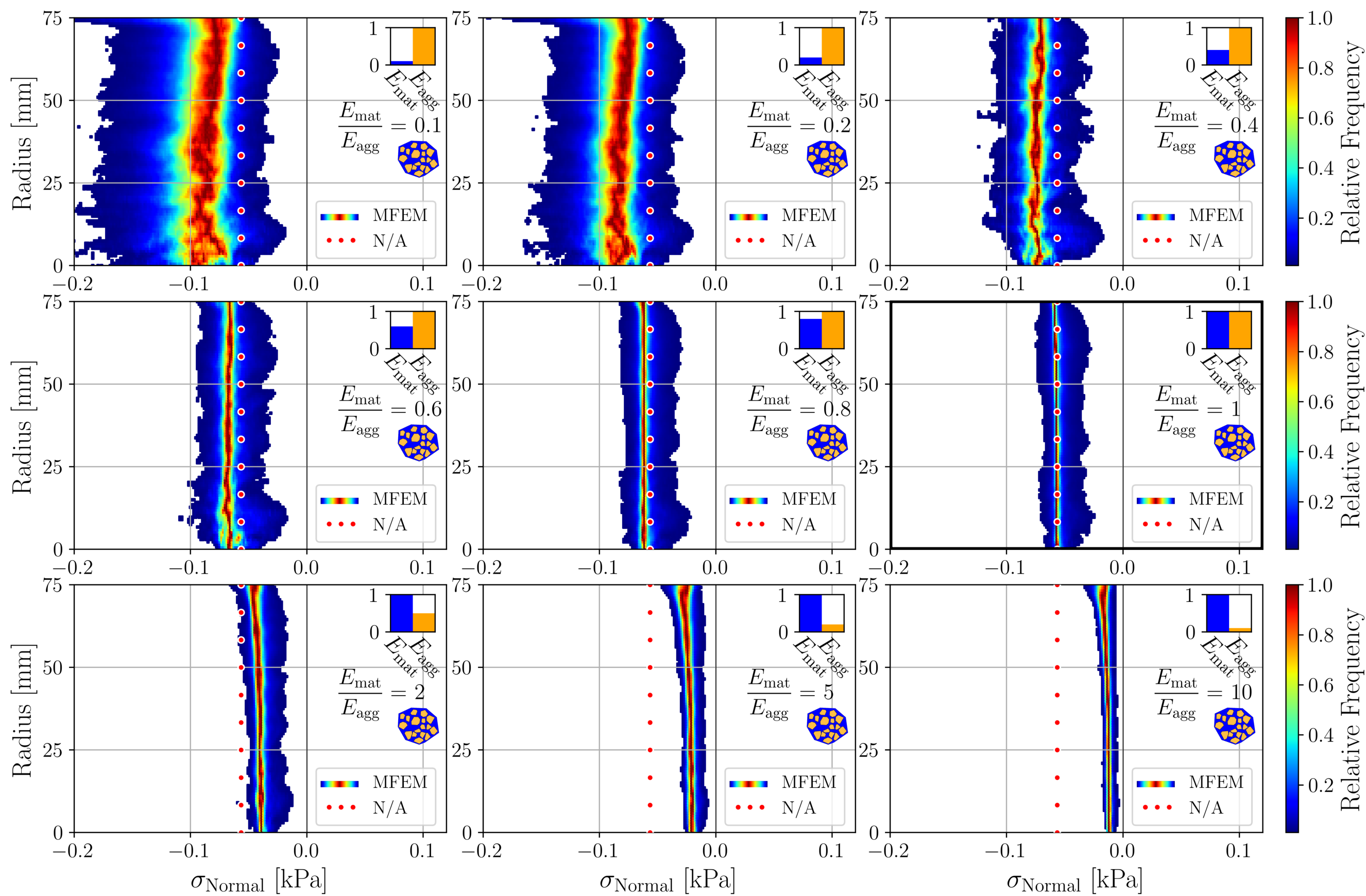}
\vspace{-3mm}
\caption{Cylinder compression: aggregates normal stress at midspan region in dependence on the heterogeneity parameter, $R_{M/A}$. Colors show relative stress frequency per horizontal layer (red = most frequent).}
\label{fig:mergedstressesaggcyl}
\end{figure*}

{\revA 
Figures~\ref{fig:bluntmergedstressesmatrix} and \ref{fig:bluntmergedstressesagg} show the stress distribution in the blunt three-point bending (TPB) model within the matrix and aggregates, respectively.
Similarly, Figures~\ref{fig:mergedstressesmatrix} and \ref{fig:mergedstressesagg} present the stress distribution for the notched TPB model, again distinguishing between matrix and aggregates.
Finally, Figures~\ref{fig:mergedstressesmatrixcyl} and \ref{fig:mergedstressesaggcyl} depict the stress distribution in the compression-loaded cylinder model within matrix and aggregates, respectively.

In all cases, stress values are visualized using a color scale normalized separately for each horizontal layer along specimen height.
The scale spans from 0 (blue, least frequent) to 1 (red, most frequent), representing the relative frequency of normal stress within each individual layer.
As such, colors indicate local stress distribution patterns
}

\clearpage

\clearpage

\begin{figure*}[b!]
    \small \hspace{7mm} a) Blunt TPB  \hspace{41mm} b) Notched TPB \hspace{34mm} c) Cylinder compression
    
    \centering
    \includegraphics[width=0.33\textwidth]{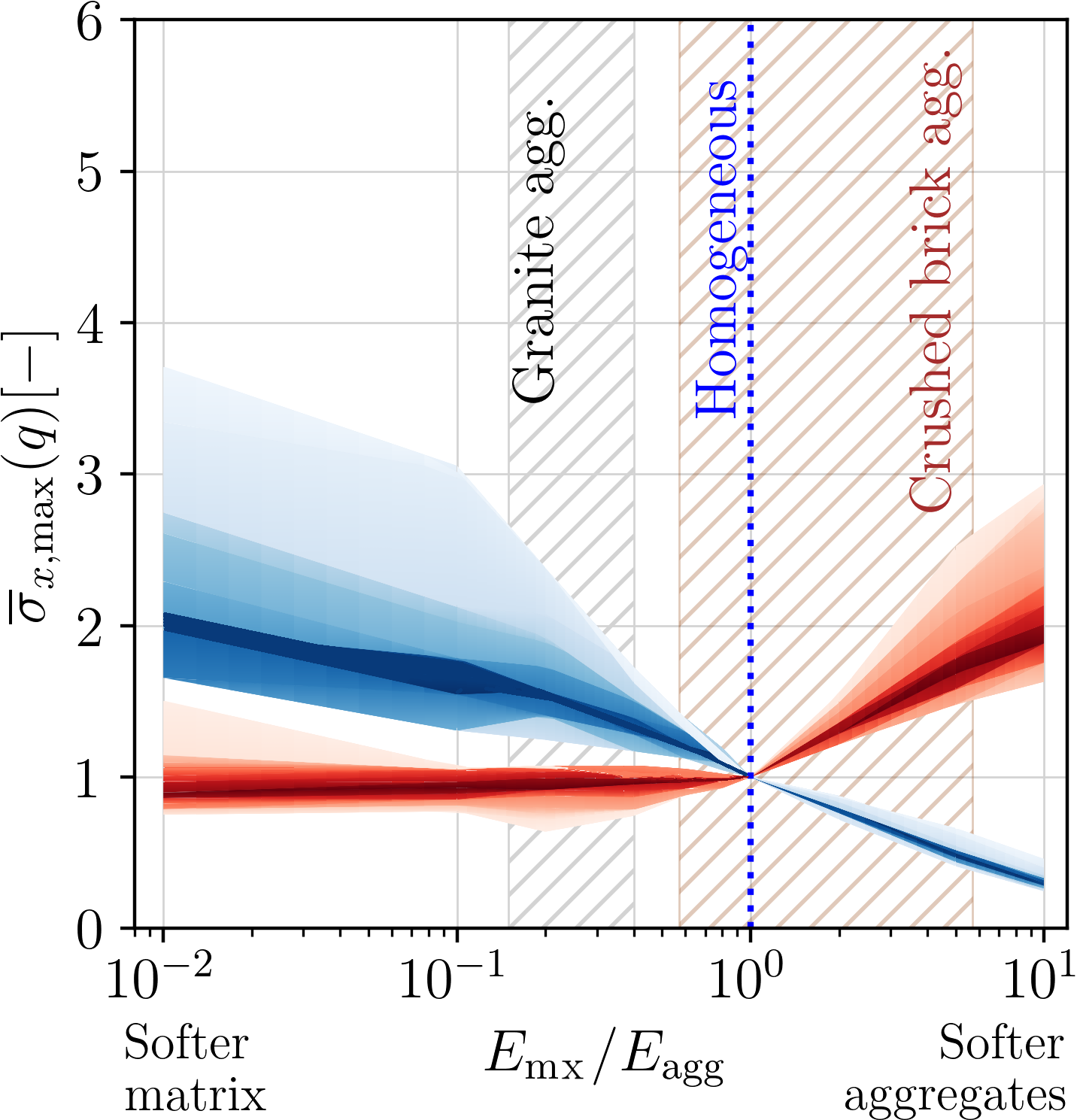}
    \includegraphics[width=0.33\textwidth]{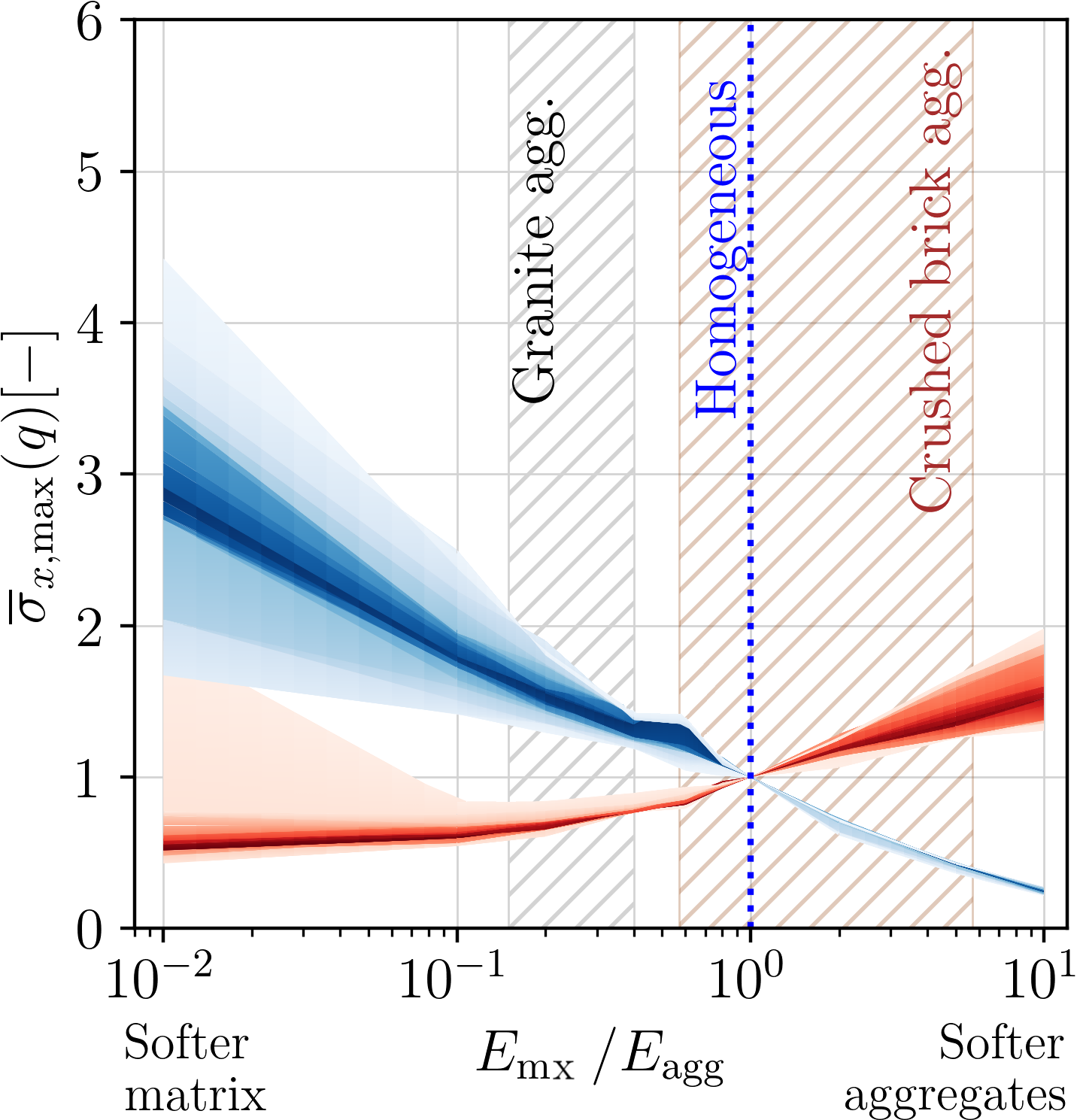}
    \includegraphics[width=0.33\textwidth]{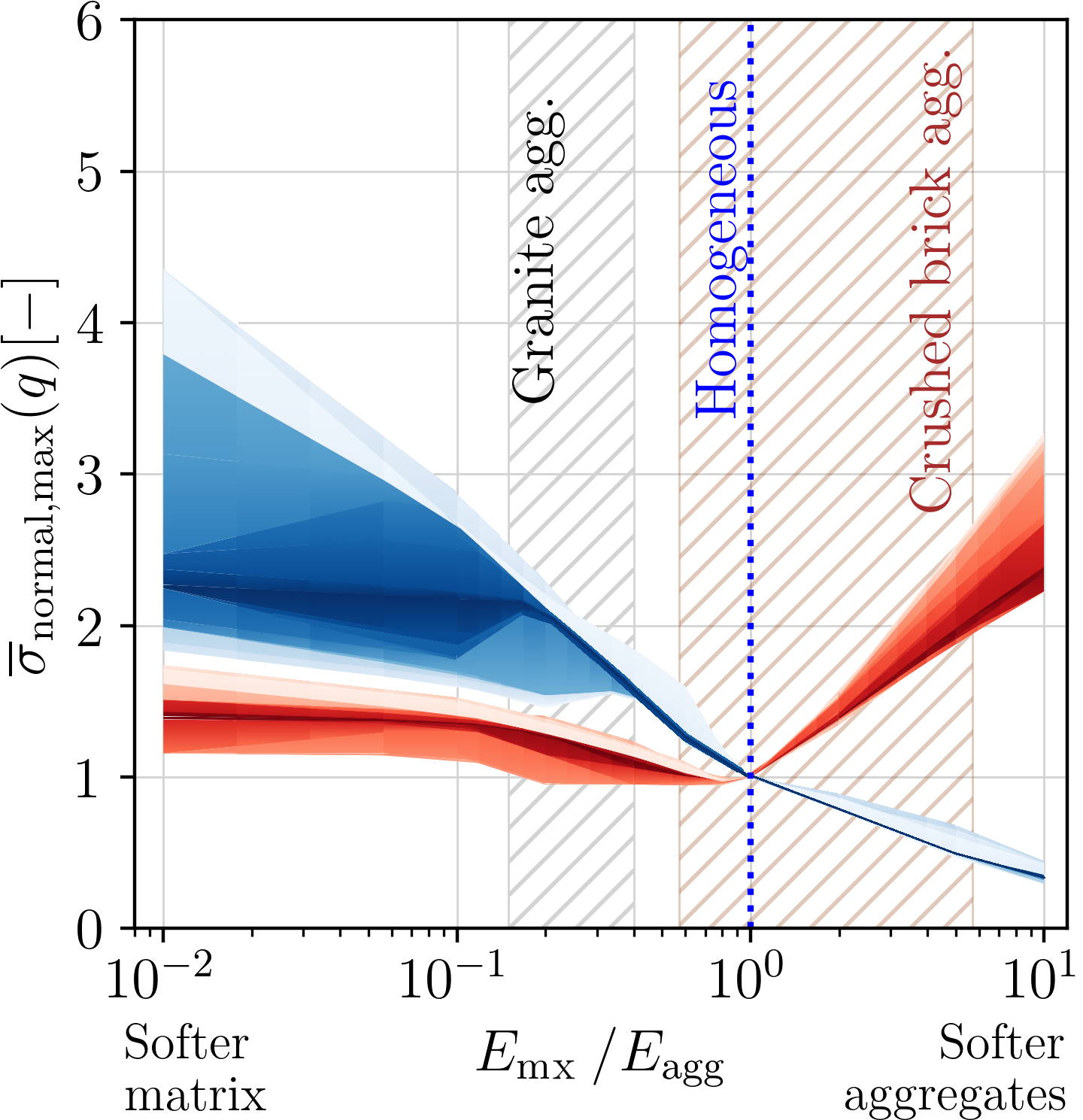}

\hspace{1mm}
  \includegraphics[width=0.65\textwidth]{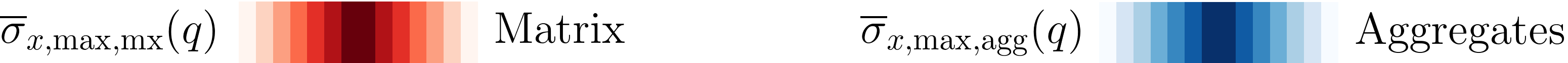}

    \caption{Normed normal stress quantiles ${\overline{\sigma}_{\mathrm{Normal},max}(q)}$.}
    \label{fig:quantiles}
\end{figure*}

This phenomenon is further illustrated in Figure~\ref{fig:quantiles}, which shows maximum normalized normal stress quantiles, \(\overline{\sigma}_{x,\max}(q)\):
\begin{equation}
    \overline{\sigma}_{x,\max}(q) = \frac{\sigma_{x,\max,\mathrm{hetero}}(q)}{\sigma_{x,\max,\mathrm{homo}}(q)},
\end{equation}
where \(\sigma_{x,\max,\mathrm{hetero}}(q)\) is the \(q\)th quantile of maximum normal stress in the heterogeneous material and \(\sigma_{x,\max,\mathrm{homo}}(q)\) is the corresponding quantile for the homogeneous material (\(R_{M/A}=1\)).

The results clearly indicate that the most favorable stress distribution occurs when the stiffness of inclusions closely matches that of the matrix.  
In this balanced state, stress is more evenly shared between phases, minimizing localized concentrations that may trigger damage.  
Deviations from this balance—either with significantly softer or stiffer inclusions—lead to pronounced stress redistribution and localization, especially at interfaces and near sharp polyhedral features.

This effect is particularly critical when replacing conventional stiff aggregates, such as granite, with alternative materials for sustainability or cost reasons.  
The mechanical consequences of high stiffness contrast must be carefully assessed, as they may compromise the composite’s structural integrity or durability.  
Our findings emphasize the importance of considering mesoscale mechanical compatibility—not just material availability—in the selection of aggregates for concrete design.

\clearpage
{\revC
\subsection{On aggregate shape influence}
To complement the study presented above, we analyze the effect of aggregate shape on stress concentrations within the mesostructure.  
The geometry of individual aggregates significantly influences the stress distribution in the composite; for a comparison of round versus polyhedral aggregates, see e.g., \cite{guo2023effect}.  
As expected, round aggregates produce lower maximal stress peaks than polyhedral ones. Even minimal deviations from roundness in otherwise nearly spherical polyhedra can trigger increased stress concentrations.

In this work, we focus exclusively on polyhedral aggregates representing crushed materials.  
We investigate the effect of increasing grain angularity (i.e., decreasing sphericity) by modifying the shrinking of tessellation cells (recall Sec.~\ref{sec:mesogeneration}). For each cell, the principal axes of inertia are determined, and the original shrinking factor $s_c$ (applied uniformly to all vertices toward the cell’s center of gravity) is scaled differently along each principal axis.

In Figure~\ref{fig:angularitystudy}a, cells are shrunk uniformly, i.e., 100\% of $s_c$ along $x_1$, $x_2$, and $x_3$. This configuration represents the most spherical polyhedral aggregates attainable with the present generation procedure, serving as the upper bound of roundness within the constraints of faceted geometry. Figure~\ref{fig:angularitystudy}b shows uneven shrinking with $x_1 = 50\%\cdot s_c$, $x_2 = 100\%\cdot s_c$, and $x_3 = 200\%\cdot s_c$, while in Figure~\ref{fig:angularitystudy}c, shrinking is more pronounced: $x_1 = 25\%\cdot s_c$, $x_2 = 100\%\cdot s_c$, and $x_3 = 250\%\cdot s_c$. Uneven shrinking generates more angular, less spherical aggregates, as illustrated in the middle panels of Figure~\ref{fig:angularitystudy}.

For each convex polyhedral aggregate, sphericity is computed using the standard Wadell formula \cite{wadell1933sphericity}:  
\begin{equation}
\Psi = \frac{\pi^{1/3} (6V)^{2/3}}{A}
\end{equation}
where $V$ is the enclosed volume and $A$ the total surface area of the aggregate. Sphericity directly quantifies deviation from a spherical form, with $\Psi = 1.0$ corresponding to a perfect sphere; lower values indicate more elongated or angular shapes.  

We also examine the angular sharpness of aggregate edges. For each face $f$, the outward unit normal $\mathbf{n}_f$ is computed. For each pair of faces $(f,g)$ sharing an edge, the dihedral angle is defined as:  
\begin{equation}
\theta_{fg} = \arccos\!\left( \mathbf{n}_f \cdot \mathbf{n}_g \right)
\end{equation}
where $\mathbf{n}_f \cdot \mathbf{n}_g$ is the dot product of the normals. The dihedral angle $\theta_{fg}$ characterizes edge sharpness: values near $0^\circ$ indicate nearly coplanar facets with no perceptible edge, $\approx 90^\circ$ corresponds to orthogonal, blocky edges, and $\approx 180^\circ$ indicates acute ridges with opposing facets.

For convex polyhedral aggregates, there exists an inherent upper bound on achievable sphericity: even in the most rounded tessellation-derived shapes, the faceted geometry prevents $\Psi$ from approaching 1.0. As a result, some degree of nonsphericity and angularity is unavoidable.

Histograms of sphericity and dihedral angles of all aggregates for each aggregate shaping variant are shown in the left panels of Figure~\ref{fig:angularitystudy}.
Finally, for a constant stiffness ratio (\(R_{M/A} =0.2\)), in a similar fashion to the stress results of Section~\ref{sec:stiffnessratiostress}, the right portion of Figure~\ref{fig:angularitystudy} shows the $\sigma_x$ stress obtained within midspan aggregates 50 unnotched three-point bending models with each grain shaping variant.

\begin{figure*}[htbp]
    \includegraphics[width=0.99\textwidth]{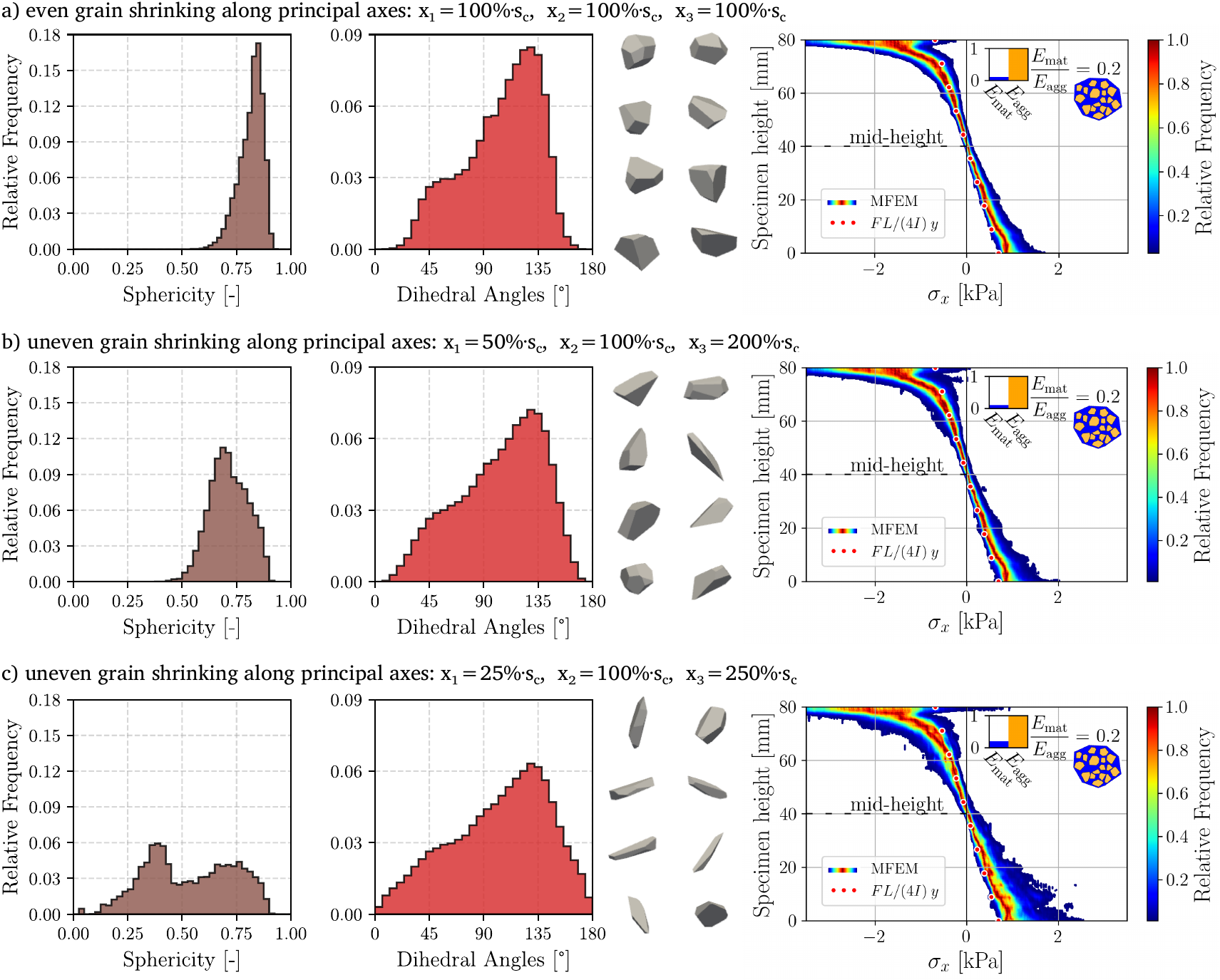}
\caption{{\revC
Effect of aggregate shaping on geometry and stress distribution. 
Left: sphericity $\Psi$ and dihedral angle $\theta_{fg}$ distributions. 
Middle: representative polyhedral aggregates showing increasing angularity from (a) to (c). 
Right: normal stress $\sigma_x$ fields, illustrating higher stress peaks with sharper aggregates.}
}
  \label{fig:angularitystudy}
  \centering
    \includegraphics[width=0.33\textwidth]{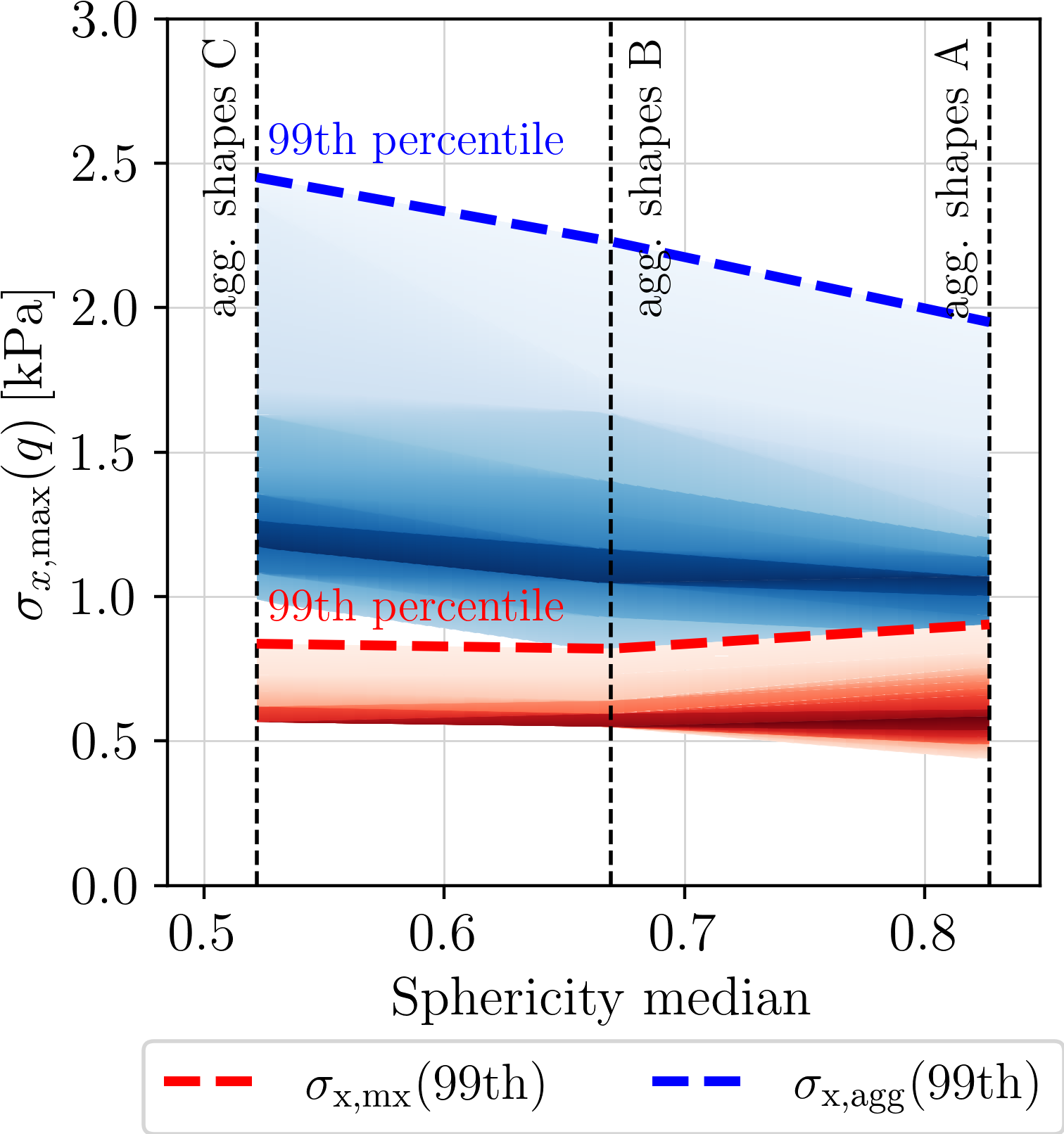}
\caption{{\revC
Distribution of maximum normal stress $\sigma_{x,\max}$ in aggregates ($\sigma_{x,agg}$) and matrix ($\sigma_{x,mx}$) for varying aggregate angularity.  
Shown are the 99th percentile peaks, highlighting that less spherical aggregates lead to increased stress concentrations. }
}
  \label{fig:angularityquantiles}
\end{figure*}

In Figure~\ref{fig:angularityquantiles}, similarly to Figure~\ref{fig:quantiles}, we now plot the distribution of normal stress within the aggregates and matrix; $\sigma_{x,\mathrm{agg}}$ and $\sigma_{x,\mathrm{mx}}$, respectively.
There, \(\sigma_{x,\max}(q)\) is the \(q\)th quantile of maximum normal stress in the material. We are now interested in the peaks: as highlighted by the 99th percentile lines form matrix and aggregates.
Clearly, less spherical aggregates tend to lead to increased stress peaks.
Given the generality of this work across materials with varying stiffness, the shown results rather indicate the general trend for already angular faceted aggregates.
The actual mechanism behind increasing stress peaks is however hidden by the interactions within each actual mesostructure, governed by mutual aggregate positions and rotations.

}
\section{\revB Conclusions}

In the presented work, a Mesoscale Finite Element Model (MFEM) of concrete was developed, enabling detailed representation of its internal heterogeneity.  
{\revA The mesostructure generation pipeline—combining weighted Voronoi tessellation, collision-checked 3D aggregate placement, and high-fidelity meshing—offers a reusable framework for simulating stress distribution and failure mechanisms in a broad class of heterogeneous materials.}  

Leveraging high-performance computing capabilities, the approach explicitly resolves the aggregate structure,  
{\revA offering transparent control over geometrical and material parameters and} offering a more transparent and versatile framework compared to existing models such as LDPM.  
{\revA This makes it suitable not only for concrete, but also for other composites where phase interaction and geometry-induced effects are critical.}  
{\revA Unlike black-box approaches such as LDPM, the MFEM setup provides physically interpretable control over input parameters, making it a viable methodology for broader applications, especially when non-traditional or recycled constituents are involved.}  

The modeling approach is particularly suitable for investigating the influence of aggregate-matrix material contrasts on stress distribution and failure mechanisms—an issue of growing relevance especially when using non-traditional aggregates such as crushed brick recyclate.  
The numerical study shows that the stiffness mismatch between the phases plays a major role in governing internal stress redistribution and damage evolution and lifetime durability. As the stiffness mismatch increases, two distinct regimes emerge:  

\begin{itemize}
    \item Stiff aggregates in a soft matrix: The compliant matrix undergoes larger deformations, allowing for stress redistribution over a wider region. This tends to reduce local stress concentrations but compromises the overall stiffness of the composite. An excessive matrix deformation may promote early microcracking, especially under fatigue or cyclic loading. 
    
    \item Soft aggregates in a stiff matrix: The stiff matrix bears the majority of the load, leading to localized stress peaks at matrix-aggregate interfaces. The aggregates no longer contribute much to stress transfer, and may act as stress concentrators rather than reinforcements. This increases the likelihood of crack initiation and propagation, often leading to brittle failure mechanisms. Ultimately, the composite resembles a matrix containing voids, with significantly impaired structural resilience.
\end{itemize}

{\revC In addition to stiffness contrast, the effect of aggregate shape was studied by varying the angularity of polyhedral aggregates. Less spherical, more angular particles were found to produce higher stress peaks, particularly within the aggregate phase, as quantified by sphericity and dihedral angle distributions. This confirms that particle geometry, alongside phase stiffness, plays a significant role in stress localization and the overall stress distribution in the composite.}

These observations underline the importance of maintaining a reasonable stiffness contrast between the phases. Ideally, aggregate materials with mechanical properties comparable to those of the matrix ensure efficient load sharing and minimal stress localization. When replacing traditional aggregates like granite with more compliant or recycled materials, such as crushed bricks, the mechanical drawbacks must be carefully assessed to avoid unintended reductions in structural integrity or durability.  

{\revA Beyond this specific case study, the developed MFEM framework offers a flexible, physically transparent tool for mesoscale stress analysis. A key methodological contribution is the histogram-based statistical post-processing of stress data across cross-sectional planes, providing quantitative insight into stress concentration distributions. This visualization technique is valuable for interpreting mesoscale mechanical behavior and can aid researchers working on heterogeneous materials modeling.}

\section{Acknowledgement}
This paper was created as part of the project No. CZ.02.01.01/00/22\_008/0004631 Materials and technologies for sustainable development within the OP JAK Program financed by the European Union and from the state budget of the Czech Republic. Numerical calculations were done within the IT4Innovations National Supercomputing Center, Czech Republic supported by the Ministry of Education, Youth and Sports of the Czech Republic through the e-INFRA CZ (ID:90254).

\section*{Data Avaiability}
The data used in this study is available at: \href{https://doi.org/10.5281/zenodo.15640474}{https://doi.org/10.5281/zenodo.15640474}.

\bibliographystyle{elsarticle-num}
\bibliography{bibliography}

\appendix
\addcontentsline{toc}{section}{Appendices}

\clearpage
\section{Notes on mesh density
\label{sec:app:meshstudy}
}

The accuracy and reliability of FEM-based numerical models are often influenced by mesh sensitivity, necessitating an appropriate mesh size to achieve both high-quality results and feasible computational efficiency.
In this study, the level of physical discretization within each model is determined by the size and placement of the aggregates.
The generated mesh consists of 3D tetrahedral elements, chosen for their compatibility with the complex geometries of the aggregates while ensuring numerical stability.
Since the analysis is purely elastic, the primary objective is to accurately capture stress distributions without accounting for nonlinearities such as cracking or plasticity. The mesh is generated using Gmsh, which provides control over element quality and ensures a well-conditioned discretization. Additionally, mesh optimization (\texttt{Mesh.Optimize}) is applied to improve element shapes and enhance numerical stability.
Mesh density is manually controlled through the maximum element size, $l_{\mathrm{max}}$, to impose sufficient refinement in the notch vicinity.

A mesh convergence study was performed to determine an optimal element size, ensuring that computed stress fields and displacement responses remained consistent across successive refinements. The final mesh configuration balances accuracy and computational cost, capturing essential features of the elastic stress distribution without excessive computational overhead.
The optimal mesh size was verified before proceeding with further numerical analysis and was subsequently adopted.
This mesh study was conducted using a three-point bending test model (see Section~\ref{sec:notchedtpb}). Several progressively finer meshes were examined, as summarized in Table~\ref{tab:elsize}.

Due to the inherent complexity of the model, the prescribed maximum element size cannot always be strictly maintained. Regions with intricate geometries or sharp features require finer elements to ensure proper mesh generation. In such cases, the meshing algorithm automatically reduces element size to maintain quality and prevent element distortion.
As a result, while the specified maximum element size serves as a guideline, local refinements occur where necessary to accommodate the model’s geometric constraints.

\begin{table}[htbp]
\small
\centering 
\begin{tabular}{l|r|r}
 $l_{\max}$  & n.o. Elements [-] & n.o. DOF [-]  \\\hline
 1\,mm & 2\,150 $\times 10^3$ &  1\,140 $\times 10^3$ \\\hline
 2\,mm & 435 $\times 10^3$ &  305 $\times 10^3$ \\\hline
 3\,mm & 232 $\times 10^3$ &  696 $\times 10^3$ \\\hline
  5\,mm & 160 $\times 10^3$ &  480 $\times 10^3$ \\\hline
  10\,mm & 158 $\times 10^3$ &  474 $\times 10^3$ \\\hline
\end{tabular}
\hspace{5mm}
\begin{minipage}{0.10\textwidth}
\centering
$1\,\mathrm{mm}$\\
\includegraphics[height=3cm,width=\textwidth,  trim=4.3cm 0cm 4.3cm 0cm, clip]{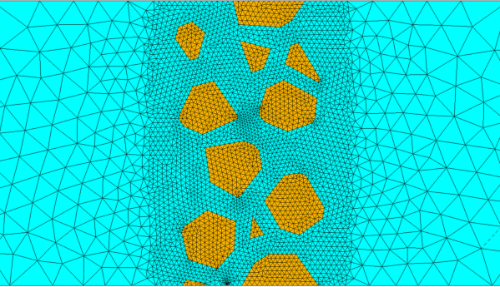}
\end{minipage}
\begin{minipage}{0.10\textwidth}
\centering
$2\,\mathrm{mm}$\\
\includegraphics[height=3cm, width=\textwidth, trim=4.3cm 0cm 4.3cm 0cm, clip]{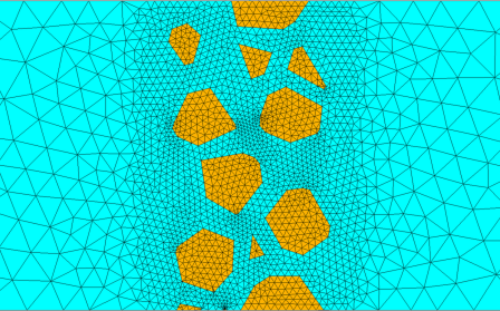}
\end{minipage}
\begin{minipage}{0.10\textwidth}
\centering
$3\,\mathrm{mm}$\\
\includegraphics[height=3cm, width=\textwidth, trim=4.3cm 0cm 4.3cm 0cm, clip]{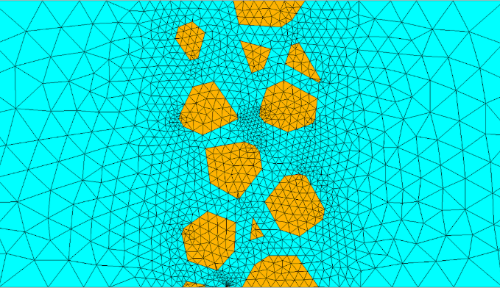}
\end{minipage}
\begin{minipage}{0.10\textwidth}
\centering
$5\,\mathrm{mm}$\\
\includegraphics[height=3cm, width=\textwidth, trim=4.3cm 0cm 4.3cm 0cm, clip]{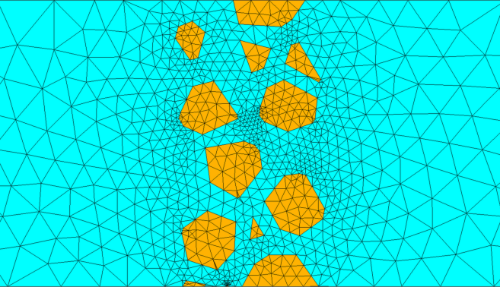}
\end{minipage}
\begin{minipage}{0.10\textwidth}
\centering
$10\,\mathrm{mm}$\\
\includegraphics[height=3cm, width=\textwidth, trim=4.3cm 0cm 4.3cm 0cm, clip]{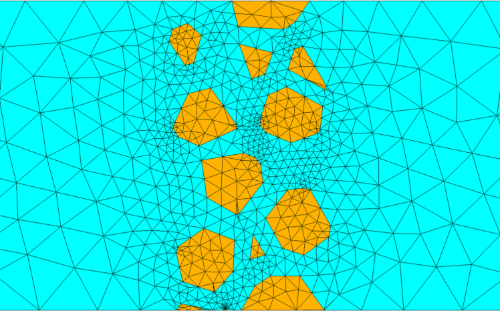}
\end{minipage}

\caption{Mesh study: midspan mesh of the TPB Model.}
\label{tab:elsize}
\end{table}

\begin{figure*}[htpb]
\centering
\includegraphics[width=0.8\linewidth]{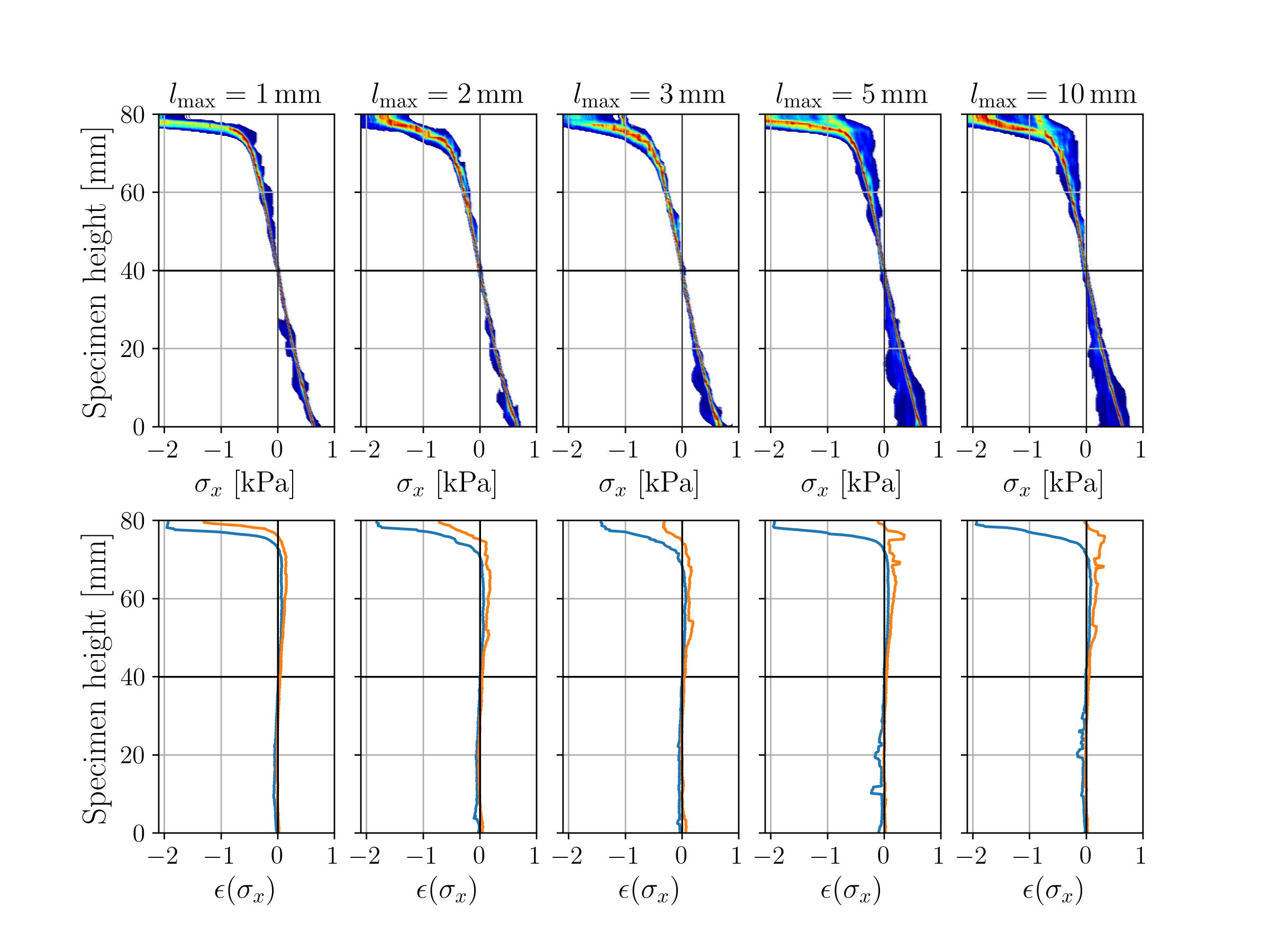}
\caption{Mesh error study}
\label{fig:meshstudy}
\end{figure*}

The study was conducted on a three-point bending beam model, and the obtained stress distribution was compared to the analytical solution.
Convergence was examined by progressively refining the mesh from coarser to finer resolutions. 
First row of Figure~\ref{fig:meshstudy} shows the respective normal stress histograms of each studied mesh.
The second row shows the absolute error, $\epsilon(\sigma_x)$, of each model envelopes respective to the analytical solution.

The results indicate that a 1 mm element size was excessively fine, leading to a substantial increase in computational cost without a proportional improvement in accuracy. Conversely, the 2 mm mesh provided sufficient precision in capturing stress distributions while maintaining a reasonable computational expense. The differences were within acceptable limits, confirming that the chosen discretization effectively captured the expected stress behavior.
Following the convergence study, a mesh with a maximum element size of 2 mm was selected for subsequent analyses.  The 2 mm mesh provided sufficient precision in capturing stress distributions while avoiding excessive computational costs associated with finer discretizations. A further reduction in element size would have significantly increased the computational burden without yielding substantial improvements in accuracy.

\clearpage
\section{ANSYS to GMSH Mesh Pipeline
\label{sec:app:gmsh}
}
\begin{lstlisting}[style=ansysstyle,caption={ANSYS to GMSH Mesh Pipeline},label={lst:ansysgmsh}]
from ansys.mapdl.core import launch_mapdl
from ansys.mapdl.reader import save_as_archive
import pyvista as pv
import os
import gmsh

# Launch MAPDL instance
mapdl = launch_mapdl(
    jobname="pyzx",
    loglevel="ERROR",
    print_com=True,
    log_apdl=True,
    override=True,
    nproc=4
)

# Load ANSYS macro containing volume geometry
mapdl.input("path_to_ansys_volume_model_macro")

# Export geometry to IGES
mapdl.cdwrite("SOLID", "modelout", "iges")

# Initialize GMSH
gmsh.initialize()
gmsh.option.setNumber("General.NumThreads", 4)  # Use 4 threads
gmsh.option.setNumber("General.Terminal", 1)    # Enable terminal output

# Import IGES geometry
v = gmsh.model.occ.importShapes("modelout.iges")  # Returns tags of imported shapes
gmsh.model.occ.synchronize()  # Synchronize GMSH CAD kernel

# Generate and optimize 3D mesh
gmsh.model.mesh.generate(3)
gmsh.model.mesh.optimize()
gmsh.write("from_gmsh.vtk")  # Export mesh for PyVista

# Extract topology and adjacency info
topology = []
adj = []
for e in gmsh.model.get_entities(3):
    dim, tag = e
    up, down = gmsh.model.getAdjacencies(dim, tag)  # 'down' = connected nodes
    elemTypes, elemTags, elemNodeTags = gmsh.model.mesh.getElements(dim, tag)
    adj.append(down)
    topology.append(elemTags)

gmsh.finalize()  # Cleanly exit GMSH

# Load mesh in PyVista, scale and resave into ANSYS *.db format
mesh = pv.read("from_gmsh.vtk")
mesh.points /= 1000.0  # Scale coordinates 

# Save mesh in MAPDL-readable archive and read into MAPDL
save_as_archive("tmp.db", mesh)
response = mapdl.cdread("db", "tmp.db")

# Merge duplicate nodes. Volumes have already been subtracted in Ansys.
# Therefore, subtracting them again in GMSH to obtain conforming mesh boundaries is unnecessary.
mapdl.allsel()
mapdl.nummrg("node", "all", 1e-7)  

# Assign aggregate material and type
mapdl.allsel() 
mapdl.emodif("all", "type", 11)
mapdl.emodif("all", "mat", 11)

# Identify largest volume for matrix
# Assign matrix material and type
mortarVolu = max(range(len(adj)), key=lambda i: len(adj[i]))
start_tag = topology[mortarVolu][0][0]
end_tag = topology[mortarVolu][0][-1]
mapdl.run(f"FLST,5,{start_tag},2,ORDE,2")
mapdl.run(f"FITEM,5,{start_tag}")
mapdl.run(f"FITEM,5,-{end_tag}")
mapdl.run("ESEL,S,,,P51X")

# Assign matrix material and type
mapdl.emodif("all", "type", 22)
mapdl.emodif("all", "mat", 22)

# Save MAPDL model
mapdl.allsel("all")  
mapdl.run("SAVE,modelname,'','dirname'")  # Replace with actual path
\end{lstlisting}

\end{document}